

\catcode`\@=11


\message{Loading jyTeX fonts...}



\font\vptrm=cmr5
\font\vptmit=cmmi5
\font\vptsy=cmsy5
\font\vptbf=cmbx5

\skewchar\vptmit='177 \skewchar\vptsy='60
\fontdimen16 \vptsy=\the\fontdimen17 \vptsy

\def\vpt{\ifmmode\err@badsizechange\else
     \@mathfontinit
     \textfont0=\vptrm  \scriptfont0=\vptrm  \scriptscriptfont0=\vptrm
     \textfont1=\vptmit \scriptfont1=\vptmit \scriptscriptfont1=\vptmit
     \textfont2=\vptsy  \scriptfont2=\vptsy  \scriptscriptfont2=\vptsy
     \textfont3=\xptex  \scriptfont3=\xptex  \scriptscriptfont3=\xptex
     \textfont\bffam=\vptbf
     \scriptfont\bffam=\vptbf
     \scriptscriptfont\bffam=\vptbf
     \@fontstyleinit
     \def\rm{\vptrm\fam=\z@}%
     \def\bf{\vptbf\fam=\bffam}%
     \def\oldstyle{\vptmit\fam=\@ne}%
     \rm\fi}


\font\viptrm=cmr6
\font\viptmit=cmmi6
\font\viptsy=cmsy6
\font\viptbf=cmbx6

\skewchar\viptmit='177 \skewchar\viptsy='60
\fontdimen16 \viptsy=\the\fontdimen17 \viptsy

\def\vipt{\ifmmode\err@badsizechange\else
     \@mathfontinit
     \textfont0=\viptrm  \scriptfont0=\vptrm  \scriptscriptfont0=\vptrm
     \textfont1=\viptmit \scriptfont1=\vptmit \scriptscriptfont1=\vptmit
     \textfont2=\viptsy  \scriptfont2=\vptsy  \scriptscriptfont2=\vptsy
     \textfont3=\xptex   \scriptfont3=\xptex  \scriptscriptfont3=\xptex
     \textfont\bffam=\viptbf
     \scriptfont\bffam=\vptbf
     \scriptscriptfont\bffam=\vptbf
     \@fontstyleinit
     \def\rm{\viptrm\fam=\z@}%
     \def\bf{\viptbf\fam=\bffam}%
     \def\oldstyle{\viptmit\fam=\@ne}%
     \rm\fi}


\font\viiptrm=cmr7
\font\viiptmit=cmmi7
\font\viiptsy=cmsy7
\font\viiptit=cmti7
\font\viiptbf=cmbx7

\skewchar\viiptmit='177 \skewchar\viiptsy='60
\fontdimen16 \viiptsy=\the\fontdimen17 \viiptsy

\def\viipt{\ifmmode\err@badsizechange\else
     \@mathfontinit
     \textfont0=\viiptrm  \scriptfont0=\vptrm  \scriptscriptfont0=\vptrm
     \textfont1=\viiptmit \scriptfont1=\vptmit \scriptscriptfont1=\vptmit
     \textfont2=\viiptsy  \scriptfont2=\vptsy  \scriptscriptfont2=\vptsy
     \textfont3=\xptex    \scriptfont3=\xptex  \scriptscriptfont3=\xptex
     \textfont\itfam=\viiptit
     \scriptfont\itfam=\viiptit
     \scriptscriptfont\itfam=\viiptit
     \textfont\bffam=\viiptbf
     \scriptfont\bffam=\vptbf
     \scriptscriptfont\bffam=\vptbf
     \@fontstyleinit
     \def\rm{\viiptrm\fam=\z@}%
     \def\it{\viiptit\fam=\itfam}%
     \def\bf{\viiptbf\fam=\bffam}%
     \def\oldstyle{\viiptmit\fam=\@ne}%
     \rm\fi}


\font\viiiptrm=cmr8
\font\viiiptmit=cmmi8
\font\viiiptsy=cmsy8
\font\viiiptit=cmti8
\font\viiiptbf=cmbx8

\skewchar\viiiptmit='177 \skewchar\viiiptsy='60
\fontdimen16 \viiiptsy=\the\fontdimen17 \viiiptsy

\def\viiipt{\ifmmode\err@badsizechange\else
     \@mathfontinit
     \textfont0=\viiiptrm  \scriptfont0=\viptrm  \scriptscriptfont0=\vptrm
     \textfont1=\viiiptmit \scriptfont1=\viptmit \scriptscriptfont1=\vptmit
     \textfont2=\viiiptsy  \scriptfont2=\viptsy  \scriptscriptfont2=\vptsy
     \textfont3=\xptex     \scriptfont3=\xptex   \scriptscriptfont3=\xptex
     \textfont\itfam=\viiiptit
     \scriptfont\itfam=\viiptit
     \scriptscriptfont\itfam=\viiptit
     \textfont\bffam=\viiiptbf
     \scriptfont\bffam=\viptbf
     \scriptscriptfont\bffam=\vptbf
     \@fontstyleinit
     \def\rm{\viiiptrm\fam=\z@}%
     \def\it{\viiiptit\fam=\itfam}%
     \def\bf{\viiiptbf\fam=\bffam}%
     \def\oldstyle{\viiiptmit\fam=\@ne}%
     \rm\fi}


\def\getixpt{%
     \font\ixptrm=cmr9
     \font\ixptmit=cmmi9
     \font\ixptsy=cmsy9
     \font\ixptit=cmti9
     \font\ixptbf=cmbx9
     \skewchar\ixptmit='177 \skewchar\ixptsy='60
     \fontdimen16 \ixptsy=\the\fontdimen17 \ixptsy}

\def\ixpt{\ifmmode\err@badsizechange\else
     \@mathfontinit
     \textfont0=\ixptrm  \scriptfont0=\viiptrm  \scriptscriptfont0=\vptrm
     \textfont1=\ixptmit \scriptfont1=\viiptmit \scriptscriptfont1=\vptmit
     \textfont2=\ixptsy  \scriptfont2=\viiptsy  \scriptscriptfont2=\vptsy
     \textfont3=\xptex   \scriptfont3=\xptex    \scriptscriptfont3=\xptex
     \textfont\itfam=\ixptit
     \scriptfont\itfam=\viiptit
     \scriptscriptfont\itfam=\viiptit
     \textfont\bffam=\ixptbf
     \scriptfont\bffam=\viiptbf
     \scriptscriptfont\bffam=\vptbf
     \@fontstyleinit
     \def\rm{\ixptrm\fam=\z@}%
     \def\it{\ixptit\fam=\itfam}%
     \def\bf{\ixptbf\fam=\bffam}%
     \def\oldstyle{\ixptmit\fam=\@ne}%
     \rm\fi}


\font\xptrm=cmr10
\font\xptmit=cmmi10
\font\xptsy=cmsy10
\font\xptex=cmex10
\font\xptit=cmti10
\font\xptsl=cmsl10
\font\xptbf=cmbx10
\font\xpttt=cmtt10
\font\xptss=cmss10
\font\xptsc=cmcsc10
\font\xptbfs=cmbx10
\font\xptbmit=cmmib10

\skewchar\xptmit='177 \skewchar\xptbmit='177 \skewchar\xptsy='60
\fontdimen16 \xptsy=\the\fontdimen17 \xptsy

\def\xpt{\ifmmode\err@badsizechange\else
     \@mathfontinit
     \textfont0=\xptrm  \scriptfont0=\viiptrm  \scriptscriptfont0=\vptrm
     \textfont1=\xptmit \scriptfont1=\viiptmit \scriptscriptfont1=\vptmit
     \textfont2=\xptsy  \scriptfont2=\viiptsy  \scriptscriptfont2=\vptsy
     \textfont3=\xptex  \scriptfont3=\xptex    \scriptscriptfont3=\xptex
     \textfont\itfam=\xptit
     \scriptfont\itfam=\viiptit
     \scriptscriptfont\itfam=\viiptit
     \textfont\bffam=\xptbf
     \scriptfont\bffam=\viiptbf
     \scriptscriptfont\bffam=\vptbf
     \textfont\bfsfam=\xptbfs
     \scriptfont\bfsfam=\viiptbf
     \scriptscriptfont\bfsfam=\vptbf
     \textfont\bmitfam=\xptbmit
     \scriptfont\bmitfam=\viiptmit
     \scriptscriptfont\bmitfam=\vptmit
     \@fontstyleinit
     \def\rm{\xptrm\fam=\z@}%
     \def\it{\xptit\fam=\itfam}%
     \def\sl{\xptsl}%
     \def\bf{\xptbf\fam=\bffam}%
     \def\tt{\xpttt}%
     \def\ss{\xptss}%
     \def\sc{\xptsc}%
     \def\bfs{\xptbfs\fam=\bfsfam}%
     \def\bmit{\fam=\bmitfam}%
     \def\oldstyle{\xptmit\fam=\@ne}%
     \rm\fi}


\def\getxipt{%
     \font\xiptrm=cmr10  scaled\magstephalf
     \font\xiptmit=cmmi10 scaled\magstephalf
     \font\xiptsy=cmsy10 scaled\magstephalf
     \font\xiptex=cmex10 scaled\magstephalf
     \font\xiptit=cmti10 scaled\magstephalf
     \font\xiptsl=cmsl10 scaled\magstephalf
     \font\xiptbf=cmbx10 scaled\magstephalf
     \font\xipttt=cmtt10 scaled\magstephalf
     \font\xiptss=cmss10 scaled\magstephalf
     \skewchar\xiptmit='177 \skewchar\xiptsy='60
     \fontdimen16 \xiptsy=\the\fontdimen17 \xiptsy}

\def\xipt{\ifmmode\err@badsizechange\else
     \@mathfontinit
     \textfont0=\xiptrm  \scriptfont0=\viiiptrm  \scriptscriptfont0=\viptrm
     \textfont1=\xiptmit \scriptfont1=\viiiptmit \scriptscriptfont1=\viptmit
     \textfont2=\xiptsy  \scriptfont2=\viiiptsy  \scriptscriptfont2=\viptsy
     \textfont3=\xiptex  \scriptfont3=\xptex     \scriptscriptfont3=\xptex
     \textfont\itfam=\xiptit
     \scriptfont\itfam=\viiiptit
     \scriptscriptfont\itfam=\viiptit
     \textfont\bffam=\xiptbf
     \scriptfont\bffam=\viiiptbf
     \scriptscriptfont\bffam=\viptbf
     \@fontstyleinit
     \def\rm{\xiptrm\fam=\z@}%
     \def\it{\xiptit\fam=\itfam}%
     \def\sl{\xiptsl}%
     \def\bf{\xiptbf\fam=\bffam}%
     \def\tt{\xipttt}%
     \def\ss{\xiptss}%
     \def\oldstyle{\xiptmit\fam=\@ne}%
     \rm\fi}


\font\xiiptrm=cmr12
\font\xiiptmit=cmmi12
\font\xiiptsy=cmsy10  scaled\magstep1
\font\xiiptex=cmex10  scaled\magstep1
\font\xiiptit=cmti12
\font\xiiptsl=cmsl12
\font\xiiptbf=cmbx12
\font\xiipttt=cmtt12
\font\xiiptss=cmss12
\font\xiiptsc=cmcsc10 scaled\magstep1
\font\xiiptbfs=cmbx10  scaled\magstep1
\font\xiiptbmit=cmmib10 scaled\magstep1

\skewchar\xiiptmit='177 \skewchar\xiiptbmit='177 \skewchar\xiiptsy='60
\fontdimen16 \xiiptsy=\the\fontdimen17 \xiiptsy

\def\xiipt{\ifmmode\err@badsizechange\else
     \@mathfontinit
     \textfont0=\xiiptrm  \scriptfont0=\viiiptrm  \scriptscriptfont0=\viptrm
     \textfont1=\xiiptmit \scriptfont1=\viiiptmit \scriptscriptfont1=\viptmit
     \textfont2=\xiiptsy  \scriptfont2=\viiiptsy  \scriptscriptfont2=\viptsy
     \textfont3=\xiiptex  \scriptfont3=\xptex     \scriptscriptfont3=\xptex
     \textfont\itfam=\xiiptit
     \scriptfont\itfam=\viiiptit
     \scriptscriptfont\itfam=\viiptit
     \textfont\bffam=\xiiptbf
     \scriptfont\bffam=\viiiptbf
     \scriptscriptfont\bffam=\viptbf
     \textfont\bfsfam=\xiiptbfs
     \scriptfont\bfsfam=\viiiptbf
     \scriptscriptfont\bfsfam=\viptbf
     \textfont\bmitfam=\xiiptbmit
     \scriptfont\bmitfam=\viiiptmit
     \scriptscriptfont\bmitfam=\viptmit
     \@fontstyleinit
     \def\rm{\xiiptrm\fam=\z@}%
     \def\it{\xiiptit\fam=\itfam}%
     \def\sl{\xiiptsl}%
     \def\bf{\xiiptbf\fam=\bffam}%
     \def\tt{\xiipttt}%
     \def\ss{\xiiptss}%
     \def\sc{\xiiptsc}%
     \def\bfs{\xiiptbfs\fam=\bfsfam}%
     \def\bmit{\fam=\bmitfam}%
     \def\oldstyle{\xiiptmit\fam=\@ne}%
     \rm\fi}


\def\getxiiipt{%
     \font\xiiiptrm=cmr12  scaled\magstephalf
     \font\xiiiptmit=cmmi12 scaled\magstephalf
     \font\xiiiptsy=cmsy9  scaled\magstep2
     \font\xiiiptit=cmti12 scaled\magstephalf
     \font\xiiiptsl=cmsl12 scaled\magstephalf
     \font\xiiiptbf=cmbx12 scaled\magstephalf
     \font\xiiipttt=cmtt12 scaled\magstephalf
     \font\xiiiptss=cmss12 scaled\magstephalf
     \skewchar\xiiiptmit='177 \skewchar\xiiiptsy='60
     \fontdimen16 \xiiiptsy=\the\fontdimen17 \xiiiptsy}

\def\xiiipt{\ifmmode\err@badsizechange\else
     \@mathfontinit
     \textfont0=\xiiiptrm  \scriptfont0=\xptrm  \scriptscriptfont0=\viiptrm
     \textfont1=\xiiiptmit \scriptfont1=\xptmit \scriptscriptfont1=\viiptmit
     \textfont2=\xiiiptsy  \scriptfont2=\xptsy  \scriptscriptfont2=\viiptsy
     \textfont3=\xivptex   \scriptfont3=\xptex  \scriptscriptfont3=\xptex
     \textfont\itfam=\xiiiptit
     \scriptfont\itfam=\xptit
     \scriptscriptfont\itfam=\viiptit
     \textfont\bffam=\xiiiptbf
     \scriptfont\bffam=\xptbf
     \scriptscriptfont\bffam=\viiptbf
     \@fontstyleinit
     \def\rm{\xiiiptrm\fam=\z@}%
     \def\it{\xiiiptit\fam=\itfam}%
     \def\sl{\xiiiptsl}%
     \def\bf{\xiiiptbf\fam=\bffam}%
     \def\tt{\xiiipttt}%
     \def\ss{\xiiiptss}%
     \def\oldstyle{\xiiiptmit\fam=\@ne}%
     \rm\fi}


\font\xivptrm=cmr12   scaled\magstep1
\font\xivptmit=cmmi12  scaled\magstep1
\font\xivptsy=cmsy10  scaled\magstep2
\font\xivptex=cmex10  scaled\magstep2
\font\xivptit=cmti12  scaled\magstep1
\font\xivptsl=cmsl12  scaled\magstep1
\font\xivptbf=cmbx12  scaled\magstep1
\font\xivpttt=cmtt12  scaled\magstep1
\font\xivptss=cmss12  scaled\magstep1
\font\xivptsc=cmcsc10 scaled\magstep2
\font\xivptbfs=cmbx10  scaled\magstep2
\font\xivptbmit=cmmib10 scaled\magstep2

\skewchar\xivptmit='177 \skewchar\xivptbmit='177 \skewchar\xivptsy='60
\fontdimen16 \xivptsy=\the\fontdimen17 \xivptsy

\def\xivpt{\ifmmode\err@badsizechange\else
     \@mathfontinit
     \textfont0=\xivptrm  \scriptfont0=\xptrm  \scriptscriptfont0=\viiptrm
     \textfont1=\xivptmit \scriptfont1=\xptmit \scriptscriptfont1=\viiptmit
     \textfont2=\xivptsy  \scriptfont2=\xptsy  \scriptscriptfont2=\viiptsy
     \textfont3=\xivptex  \scriptfont3=\xptex  \scriptscriptfont3=\xptex
     \textfont\itfam=\xivptit
     \scriptfont\itfam=\xptit
     \scriptscriptfont\itfam=\viiptit
     \textfont\bffam=\xivptbf
     \scriptfont\bffam=\xptbf
     \scriptscriptfont\bffam=\viiptbf
     \textfont\bfsfam=\xivptbfs
     \scriptfont\bfsfam=\xptbfs
     \scriptscriptfont\bfsfam=\viiptbf
     \textfont\bmitfam=\xivptbmit
     \scriptfont\bmitfam=\xptbmit
     \scriptscriptfont\bmitfam=\viiptmit
     \@fontstyleinit
     \def\rm{\xivptrm\fam=\z@}%
     \def\it{\xivptit\fam=\itfam}%
     \def\sl{\xivptsl}%
     \def\bf{\xivptbf\fam=\bffam}%
     \def\tt{\xivpttt}%
     \def\ss{\xivptss}%
     \def\sc{\xivptsc}%
     \def\bfs{\xivptbfs\fam=\bfsfam}%
     \def\bmit{\fam=\bmitfam}%
     \def\oldstyle{\xivptmit\fam=\@ne}%
     \rm\fi}


\font\xviiptrm=cmr17
\font\xviiptmit=cmmi12 scaled\magstep2
\font\xviiptsy=cmsy10 scaled\magstep3
\font\xviiptex=cmex10 scaled\magstep3
\font\xviiptit=cmti12 scaled\magstep2
\font\xviiptbf=cmbx12 scaled\magstep2
\font\xviiptbfs=cmbx10 scaled\magstep3

\skewchar\xviiptmit='177 \skewchar\xviiptsy='60
\fontdimen16 \xviiptsy=\the\fontdimen17 \xviiptsy

\def\xviipt{\ifmmode\err@badsizechange\else
     \@mathfontinit
     \textfont0=\xviiptrm  \scriptfont0=\xiiptrm  \scriptscriptfont0=\viiiptrm
     \textfont1=\xviiptmit \scriptfont1=\xiiptmit \scriptscriptfont1=\viiiptmit
     \textfont2=\xviiptsy  \scriptfont2=\xiiptsy  \scriptscriptfont2=\viiiptsy
     \textfont3=\xviiptex  \scriptfont3=\xiiptex  \scriptscriptfont3=\xptex
     \textfont\itfam=\xviiptit
     \scriptfont\itfam=\xiiptit
     \scriptscriptfont\itfam=\viiiptit
     \textfont\bffam=\xviiptbf
     \scriptfont\bffam=\xiiptbf
     \scriptscriptfont\bffam=\viiiptbf
     \textfont\bfsfam=\xviiptbfs
     \scriptfont\bfsfam=\xiiptbfs
     \scriptscriptfont\bfsfam=\viiiptbf
     \@fontstyleinit
     \def\rm{\xviiptrm\fam=\z@}%
     \def\it{\xviiptit\fam=\itfam}%
     \def\bf{\xviiptbf\fam=\bffam}%
     \def\bfs{\xviiptbfs\fam=\bfsfam}%
     \def\oldstyle{\xviiptmit\fam=\@ne}%
     \rm\fi}


\font\xxiptrm=cmr17  scaled\magstep1


\def\xxipt{\ifmmode\err@badsizechange\else
     \@mathfontinit
     \@fontstyleinit
     \def\rm{\xxiptrm\fam=\z@}%
     \rm\fi}


\font\xxvptrm=cmr17  scaled\magstep2


\def\xxvpt{\ifmmode\err@badsizechange\else
     \@mathfontinit
     \@fontstyleinit
     \def\rm{\xxvptrm\fam=\z@}%
     \rm\fi}




\message{Loading jyTeX macros...}

\message{modifications to plain.tex,}


\def\newcount{\alloc@0\count\countdef\insc@unt}
\def\newdimen{\alloc@1\dimen\dimendef\insc@unt}
\def\newskip{\alloc@2\skip\skipdef\insc@unt}
\def\newmuskip{\alloc@3\muskip\muskipdef\@cclvi}
\def\newbox{\alloc@4\box\chardef\insc@unt}
\def\newtoks{\alloc@5\toks\toksdef\@cclvi}
\def\newhelp#1#2{\newtoks#1\global#1\expandafter{\csname#2\endcsname}}
\def\newread{\alloc@6\read\chardef\sixt@@n}
\def\newwrite{\alloc@7\write\chardef\sixt@@n}
\def\newfam{\alloc@8\fam\chardef\sixt@@n}
\def\newinsert#1{\global\advance\insc@unt by\m@ne
     \ch@ck0\insc@unt\count
     \ch@ck1\insc@unt\dimen
     \ch@ck2\insc@unt\skip
     \ch@ck4\insc@unt\box
     \allocationnumber=\insc@unt
     \global\chardef#1=\allocationnumber
     \wlog{\string#1=\string\insert\the\allocationnumber}}
\def\newif#1{\count@\escapechar \escapechar\m@ne
     \expandafter\expandafter\expandafter
          \xdef\@if#1{true}{\let\noexpand#1=\noexpand\iftrue}%
     \expandafter\expandafter\expandafter
          \xdef\@if#1{false}{\let\noexpand#1=\noexpand\iffalse}%
     \global\@if#1{false}\escapechar=\count@}


\newlinechar=`\^^J
\overfullrule=0pt




\let\itfam=\undefined

\let\bffam=\undefined

\count18=3


\chardef\sharps="19


\mathchardef\alpha="710B
\mathchardef\beta="710C
\mathchardef\gamma="710D
\mathchardef\delta="710E
\mathchardef\epsilon="710F
\mathchardef\zeta="7110
\mathchardef\eta="7111
\mathchardef\theta="7112
\mathchardef\iota="7113
\mathchardef\kappa="7114
\mathchardef\lambda="7115
\mathchardef\mu="7116
\mathchardef\nu="7117
\mathchardef\xi="7118
\mathchardef\pi="7119
\mathchardef\rho="711A
\mathchardef\sigma="711B
\mathchardef\tau="711C
\mathchardef\upsilon="711D
\mathchardef\phi="711E
\mathchardef\chi="711F
\mathchardef\psi="7120
\mathchardef\omega="7121
\mathchardef\varepsilon="7122
\mathchardef\vartheta="7123
\mathchardef\varpi="7124
\mathchardef\varrho="7125
\mathchardef\varsigma="7126
\mathchardef\varphi="7127
\mathchardef\imath="717B
\mathchardef\jmath="717C
\mathchardef\ell="7160
\mathchardef\wp="717D
\mathchardef\partial="7140
\mathchardef\flat="715B
\mathchardef\natural="715C
\mathchardef\sharp="715D



\def\angle{{\vbox{\ialign{$\m@th\scriptstyle##$\crcr
     \not\mathrel{\mkern14mu}\crcr
     \noalign{\nointerlineskip}
     \mkern2.5mu\leaders\hrule height.34\rp@\hfill\mkern2.5mu\crcr}}}}
\def\vdots{\vbox{\baselineskip4\rp@ \lineskiplimit\z@
     \kern6\rp@\hbox{.}\hbox{.}\hbox{.}}}
\def\ddots{\mathinner{\mkern1mu\raise7\rp@\vbox{\kern7\rp@\hbox{.}}\mkern2mu
     \raise4\rp@\hbox{.}\mkern2mu\raise\rp@\hbox{.}\mkern1mu}}
\def\overrightarrow#1{\vbox{\ialign{##\crcr
     \rightarrowfill\crcr
     \noalign{\kern-\rp@\nointerlineskip}
     $\hfil\displaystyle{#1}\hfil$\crcr}}}
\def\overleftarrow#1{\vbox{\ialign{##\crcr
     \leftarrowfill\crcr
     \noalign{\kern-\rp@\nointerlineskip}
     $\hfil\displaystyle{#1}\hfil$\crcr}}}
\def\overbrace#1{\mathop{\vbox{\ialign{##\crcr
     \noalign{\kern3\rp@}
     \downbracefill\crcr
     \noalign{\kern3\rp@\nointerlineskip}
     $\hfil\displaystyle{#1}\hfil$\crcr}}}\limits}
\def\underbrace#1{\mathop{\vtop{\ialign{##\crcr
     $\hfil\displaystyle{#1}\hfil$\crcr
     \noalign{\kern3\rp@\nointerlineskip}
     \upbracefill\crcr
     \noalign{\kern3\rp@}}}}\limits}
\def\big#1{{\hbox{$\left#1\vbox to8.5\rp@ {}\right.\n@space$}}}
\def\Big#1{{\hbox{$\left#1\vbox to11.5\rp@ {}\right.\n@space$}}}
\def\bigg#1{{\hbox{$\left#1\vbox to14.5\rp@ {}\right.\n@space$}}}
\def\Bigg#1{{\hbox{$\left#1\vbox to17.5\rp@ {}\right.\n@space$}}}
\def\@vereq#1#2{\lower.5\rp@\vbox{\baselineskip\z@skip\lineskip-.5\rp@
     \ialign{$\m@th#1\hfil##\hfil$\crcr#2\crcr=\crcr}}}
\def\rlh@#1{\vcenter{\hbox{\ooalign{\raise2\rp@
     \hbox{$#1\rightharpoonup$}\crcr
     $#1\leftharpoondown$}}}}
\def\bordermatrix#1{\begingroup\m@th
     \setbox\z@\vbox{%
          \def\cr{\crcr\noalign{\kern2\rp@\global\let\cr\endline}}%
          \ialign{$##$\hfil\kern2\rp@\kern\p@renwd
               &\thinspace\hfil$##$\hfil&&\quad\hfil$##$\hfil\crcr
               \omit\strut\hfil\crcr
               \noalign{\kern-\baselineskip}%
               #1\crcr\omit\strut\cr}}%
     \setbox\tw@\vbox{\unvcopy\z@\global\setbox\@ne\lastbox}%
     \setbox\tw@\hbox{\unhbox\@ne\unskip\global\setbox\@ne\lastbox}%
     \setbox\tw@\hbox{$\kern\wd\@ne\kern-\p@renwd\left(\kern-\wd\@ne
          \global\setbox\@ne\vbox{\box\@ne\kern2\rp@}%
          \vcenter{\kern-\ht\@ne\unvbox\z@\kern-\baselineskip}%
          \,\right)$}%
     \null\;\vbox{\kern\ht\@ne\box\tw@}\endgroup}
\def\endinsert{\egroup
     \if@mid\dimen@\ht\z@
          \advance\dimen@\dp\z@
          \advance\dimen@12\rp@
          \advance\dimen@\pagetotal
          \ifdim\dimen@>\pagegoal\@midfalse\p@gefalse\fi
     \fi
     \if@mid\bigskip\box\z@
          \bigbreak
     \else\insert\topins{\penalty100 \splittopskip\z@skip
               \splitmaxdepth\maxdimen\floatingpenalty\z@
               \ifp@ge\dimen@\dp\z@
                    \vbox to\vsize{\unvbox\z@\kern-\dimen@}%
               \else\box\z@\nobreak\bigskip
               \fi}%
     \fi
     \endgroup}


\def\cases#1{\left\{\,\vcenter{\m@th
     \ialign{$##\hfil$&\quad##\hfil\crcr#1\crcr}}\right.}
\def\matrix#1{\null\,\vcenter{\m@th
     \ialign{\hfil$##$\hfil&&\quad\hfil$##$\hfil\crcr
          \mathstrut\crcr
          \noalign{\kern-\baselineskip}
          #1\crcr
          \mathstrut\crcr
          \noalign{\kern-\baselineskip}}}\,}


\newif\ifraggedbottom

\def\raggedbottom{\ifraggedbottom\else
     \advance\topskip by\z@ plus60pt \raggedbottomtrue\fi}%
\def\normalbottom{\ifraggedbottom
     \advance\topskip by\z@ plus-60pt \raggedbottomfalse\fi}

\message{hacks,}


\toksdef\toks@i=1
\toksdef\toks@ii=2


\def\TeX{T\kern-.1667em \lower.5ex \hbox{E}\kern-.125em X\null}
\def\jyTeX{{\leavevmode
     \raise.587ex \hbox{\it\j}\kern-.1em \lower.048ex \hbox{\it y}\kern-.12em
     \TeX}}

\let\then=\iftrue
\def\ifnoarg#1\then{\def\hack@{#1}\ifx\hack@\empty}
\def\ifundefined#1\then{%
     \expandafter\ifx\csname\expandafter\blank\string#1\endcsname\relax}
\def\useif#1\then{\csname#1\endcsname}
\def\usename#1{\csname#1\endcsname}
\def\useafter#1#2{\expandafter#1\csname#2\endcsname}

\long\def\loop#1\repeat{\def\@iterate{#1\expandafter\@iterate\fi}\@iterate
     \let\@iterate=\relax}

\let\TeXend=\end
\def\begin#1{\begingroup\def\@@blockname{#1}\usename{begin#1}}
\def\end#1{\usename{end#1}\def\hack@{#1}%
     \ifx\@@blockname\hack@
          \endgroup
     \else\err@badgroup\hack@\@@blockname
     \fi}
\def\@@blockname{}

\def\defaultoption[#1]#2{%
     \def\hack@{\ifx\hack@ii[\toks@={#2}\else\toks@={#2[#1]}\fi\the\toks@}%
     \futurelet\hack@ii\hack@}

\def\markup#1{\let\@@marksf=\empty
     \ifhmode\edef\@@marksf{\spacefactor=\the\spacefactor\relax}\/\fi
     ${}^{\hbox{\subscriptfonts#1}}$\@@marksf}


\newtoks\shortyear
\newtoks\militaryhour
\newtoks\standardhour
\newtoks\minute
\newtoks\amorpm

\def\settime{\count@=\time\divide\count@ by60
     \militaryhour=\expandafter{\number\count@}%
     {\multiply\count@ by-60 \advance\count@ by\time
          \xdef\hack@{\ifnum\count@<10 0\fi\number\count@}}%
     \minute=\expandafter{\hack@}%
     \ifnum\count@<12
          \amorpm={am}
     \else\amorpm={pm}
          \ifnum\count@>12 \advance\count@ by-12 \fi
     \fi
     \standardhour=\expandafter{\number\count@}%
     \def\hack@19##1##2{\shortyear={##1##2}}%
          \expandafter\hack@\the\year}

\def\monthword#1{%
     \ifcase#1
          $\bullet$\err@badcountervalue{monthword}%
          \or January\or February\or March\or April\or May\or June%
          \or July\or August\or September\or October\or November\or December%
     \else$\bullet$\err@badcountervalue{monthword}%
     \fi}

\def\monthabbr#1{%
     \ifcase#1
          $\bullet$\err@badcountervalue{monthabbr}%
          \or Jan\or Feb\or Mar\or Apr\or May\or Jun%
          \or Jul\or Aug\or Sep\or Oct\or Nov\or Dec%
     \else$\bullet$\err@badcountervalue{monthabbr}%
     \fi}

\def\militarytime{\the\militaryhour:\the\minute}
\def\standardtime{\the\standardhour:\the\minute}


\def\@setnumstyle#1#2{\expandafter\global\expandafter\expandafter
     \expandafter\let\expandafter\expandafter
     \csname @\expandafter\blank\string#1style\endcsname
     \csname#2\endcsname}
\def\numstyle#1{\usename{@\expandafter\blank\string#1style}#1}
\def\ifblank#1\then{\useafter\ifx{@\expandafter\blank\string#1}\blank}

\def\blank#1{}

\def\Roman#1{\expandafter\uppercase\expandafter{\romannumeral#1}}
\def\alphabetic#1{%
     \ifcase#1
          $\bullet$\err@badcountervalue{alphabetic}%
          \or a\or b\or c\or d\or e\or f\or g\or h\or i\or j\or k\or l\or m%
          \or n\or o\or p\or q\or r\or s\or t\or u\or v\or w\or x\or y\or z%
     \else$\bullet$\err@badcountervalue{alphabetic}%
     \fi}
\def\Alphabetic#1{\expandafter\uppercase\expandafter{\alphabetic{#1}}}
\def\symbols#1{%
     \ifcase#1
          $\bullet$\err@badcountervalue{symbols}%
          \or*\or\dag\or\ddag\or\S\or$\|$%
          \or**\or\dag\dag\or\ddag\ddag\or\S\S\or$\|\|$%
     \else$\bullet$\err@badcountervalue{symbols}%
     \fi}


\catcode`\^^?=13 \def^^?{\relax}

\def\trimleading#1\to#2{\edef#2{#1}%
     \expandafter\@trimleading\expandafter#2#2^^?^^?}
\def\@trimleading#1#2#3^^?{\ifx#2^^?\def#1{}\else\def#1{#2#3}\fi}

\def\trimtrailing#1\to#2{\edef#2{#1}%
     \expandafter\@trimtrailing\expandafter#2#2^^? ^^?\relax}
\def\@trimtrailing#1#2 ^^?#3{\ifx#3\relax\toks@={}%
     \else\def#1{#2}\toks@={\trimtrailing#1\to#1}\fi
     \the\toks@}

\def\trim#1\to#2{\trimleading#1\to#2\trimtrailing#2\to#2}

\catcode`\^^?=15


\long\def\additemL#1\to#2{\toks@={\^^\{#1}}\toks@ii=\expandafter{#2}%
     \xdef#2{\the\toks@\the\toks@ii}}

\long\def\additemR#1\to#2{\toks@={\^^\{#1}}\toks@ii=\expandafter{#2}%
     \xdef#2{\the\toks@ii\the\toks@}}

\def\getitemL#1\to#2{\expandafter\@getitemL#1\hack@#1#2}
\def\@getitemL\^^\#1#2\hack@#3#4{\def#4{#1}\def#3{#2}}

\message{font macros,}


\newdimen\rp@
\newcount\@@sizeindex \@@sizeindex=0
\newcount\@@factori
\newcount\@@factorii
\newcount\@@factoriii
\newcount\@@factoriv

\countdef\maxfam=18
\newfam\itfam
\newfam\bffam
\newfam\bfsfam
\newfam\bmitfam

\def\@mathfontinit{\count@=4
     \loop\textfont\count@=\nullfont
          \scriptfont\count@=\nullfont
          \scriptscriptfont\count@=\nullfont
          \ifnum\count@<\maxfam\advance\count@ by\@ne
     \repeat}

\def\@fontstyleinit{%
     \def\it{\err@fontnotavailable\it}%
     \def\bf{\err@fontnotavailable\bf}%
     \def\bfs{\err@bfstobf}%
     \def\bmit{\err@fontnotavailable\bmit}%
     \def\sc{\err@fontnotavailable\sc}%
     \def\sl{\err@sltoit}%
     \def\ss{\err@fontnotavailable\ss}%
     \def\tt{\err@fontnotavailable\tt}}

\def\@parameterinit#1{\rm\rp@=.1em \@getscaling{#1}%
     \let\^^\=\@doscaling\scalingskipslist
     \setbox\strutbox=\hbox{\vrule
          height.708\baselineskip depth.292\baselineskip width\z@}}

\def\@getfactor#1#2#3#4{\@@factori=#1 \@@factorii=#2
     \@@factoriii=#3 \@@factoriv=#4}

\def\@getscaling#1{\count@=#1 \advance\count@ by-\@@sizeindex\@@sizeindex=#1
     \ifnum\count@<0
          \let\@mulordiv=\divide
          \let\@divormul=\multiply
          \multiply\count@ by\m@ne
     \else\let\@mulordiv=\multiply
          \let\@divormul=\divide
     \fi
     \edef\@@scratcha{\ifcase\count@                {1}{1}{1}{1}\or
          {1}{7}{23}{3}\or     {2}{5}{3}{1}\or      {9}{89}{13}{1}\or
          {6}{25}{6}{1}\or     {8}{71}{14}{1}\or    {6}{25}{36}{5}\or
          {1}{7}{53}{4}\or     {12}{125}{108}{5}\or {3}{14}{53}{5}\or
          {6}{41}{17}{1}\or    {13}{31}{13}{2}\or   {9}{107}{71}{2}\or
          {11}{139}{124}{3}\or {1}{6}{43}{2}\or     {10}{107}{42}{1}\or
          {1}{5}{43}{2}\or     {5}{69}{65}{1}\or    {11}{97}{91}{2}\fi}%
     \expandafter\@getfactor\@@scratcha}

\def\@doscaling#1{\@mulordiv#1by\@@factori\@divormul#1by\@@factorii
     \@mulordiv#1by\@@factoriii\@divormul#1by\@@factoriv}


\newskip\headskip
\newskip\footskip

\def\typesize=#1pt{\count@=#1 \advance\count@ by-10
     \ifcase\count@
          \@setsizex\or\err@badtypesize\or
          \@setsizexii\or\err@badtypesize\or
          \@setsizexiv
     \else\err@badtypesize
     \fi}

\def\@setsizex{\getixpt
     \def\subsubscriptfonts{\vpt}%
          \def\subsubscriptsize{\vpt\@parameterinit{-8}}%
     \def\subscriptfonts{\viipt}\def\subscriptsize{\viipt\@parameterinit{-4}}%
     \def\footnotefonts{\viiipt}\def\footnotesize{\viiipt\@parameterinit{-2}}%
     \def\smallfonts{\ixpt}\def\smallsize{\ixpt\@parameterinit{-1}}%
     \def\normalfonts{\xpt}\def\normalsize{\xpt\@parameterinit{0}}%
     \def\bigfonts{\xiipt}\def\bigsize{\xiipt\@parameterinit{2}}%
     \def\Bigfonts{\xivpt}\def\Bigsize{\xivpt\@parameterinit{4}}%
     \def\biggfonts{\xviipt}\def\biggsize{\xviipt\@parameterinit{6}}%
     \def\Biggfonts{\xxipt}\def\Biggsize{\xxipt\@parameterinit{8}}%
     \def\tinyfonts{\vpt}\def\tinysize{\vpt\@parameterinit{-8}}%
     \def\HUGEFONTS{\xxvpt}\def\HUGESIZE{\xxvpt\@parameterinit{10}}%
     \normalsize\fixedskipslist}

\def\@setsizexii{\getxipt
     \def\subsubscriptfonts{\vipt}%
          \def\subsubscriptsize{\vipt\@parameterinit{-6}}%
     \def\subscriptfonts{\viiipt}%
          \def\subscriptsize{\viiipt\@parameterinit{-2}}%
     \def\footnotefonts{\xpt}\def\footnotesize{\xpt\@parameterinit{0}}%
     \def\smallfonts{\xipt}\def\smallsize{\xipt\@parameterinit{1}}%
     \def\normalfonts{\xiipt}\def\normalsize{\xiipt\@parameterinit{2}}%
     \def\bigfonts{\xivpt}\def\bigsize{\xivpt\@parameterinit{4}}%
     \def\Bigfonts{\xviipt}\def\Bigsize{\xviipt\@parameterinit{6}}%
     \def\biggfonts{\xxipt}\def\biggsize{\xxipt\@parameterinit{8}}%
     \def\Biggfonts{\xxvpt}\def\Biggsize{\xxvpt\@parameterinit{10}}%
     \def\tinyfonts{\vpt}\def\tinysize{\vpt\@parameterinit{-8}}%
     \def\HUGEFONTS{\xxvpt}\def\HUGESIZE{\xxvpt\@parameterinit{10}}%
     \normalsize\fixedskipslist}

\def\@setsizexiv{\getxiiipt
     \def\subsubscriptfonts{\viipt}%
          \def\subsubscriptsize{\viipt\@parameterinit{-4}}%
     \def\subscriptfonts{\xpt}\def\subscriptsize{\xpt\@parameterinit{0}}%
     \def\footnotefonts{\xiipt}\def\footnotesize{\xiipt\@parameterinit{2}}%
     \def\smallfonts{\xiiipt}\def\smallsize{\xiiipt\@parameterinit{3}}%
     \def\normalfonts{\xivpt}\def\normalsize{\xivpt\@parameterinit{4}}%
     \def\bigfonts{\xviipt}\def\bigsize{\xviipt\@parameterinit{6}}%
     \def\Bigfonts{\xxipt}\def\Bigsize{\xxipt\@parameterinit{8}}%
     \def\biggfonts{\xxvpt}\def\biggsize{\xxvpt\@parameterinit{10}}%
     \def\Biggfonts{\err@sizetoolarge\Biggfonts\HUGEFONTS}%
          \def\Biggsize{\err@sizetoolarge\Biggsize\HUGESIZE}%
     \def\tinyfonts{\vpt}\def\tinysize{\vpt\@parameterinit{-8}}%
     \def\HUGEFONTS{\xxvpt}\def\HUGESIZE{\xxvpt\@parameterinit{10}}%
     \normalsize\fixedskipslist}

\def\subsubscriptfonts{\vpt} \def\subsubscriptsize{\vpt\@parameterinit{-8}}
\def\subscriptfonts{\viipt}  \def\subscriptsize{\viipt\@parameterinit{-4}}
\def\footnotefonts{\viiipt}  \def\footnotesize{\viiipt\@parameterinit{-2}}
\def\smallfonts{\err@sizenotavailable\smallfonts}
                             \def\smallsize{\ixpt\@parameterinit{-1}}
\def\normalfonts{\xpt}       \def\normalsize{\xpt\@parameterinit{0}}
\def\bigfonts{\xiipt}        \def\bigsize{\xiipt\@parameterinit{2}}
\def\Bigfonts{\xivpt}        \def\Bigsize{\xivpt\@parameterinit{4}}
\def\biggfonts{\xviipt}      \def\biggsize{\xviipt\@parameterinit{6}}
\def\Biggfonts{\xxipt}       \def\Biggsize{\xxipt\@parameterinit{8}}
\def\tinyfonts{\vpt}         \def\tinysize{\vpt\@parameterinit{-8}}
\def\HUGEFONTS{\xxvpt}       \def\HUGESIZE{\xxvpt\@parameterinit{10}}

\message{document layout,}


\newtoks\everyoutput \everyoutput={}
\newdimen\depthofpage
\newcount\pagenum \pagenum=0

\newdimen\oddtopmargin  \newdimen\eventopmargin
\newdimen\oddleftmargin \newdimen\evenleftmargin
\newtoks\oddhead        \newtoks\evenhead
\newtoks\oddfoot        \newtoks\evenfoot

\def\topmargin{\afterassignment\@seteventop\oddtopmargin}
\def\leftmargin{\afterassignment\@setevenleft\oddleftmargin}
\def\head{\afterassignment\@setevenhead\oddhead}
\def\foot{\afterassignment\@setevenfoot\oddfoot}

\def\@seteventop{\eventopmargin=\oddtopmargin}
\def\@setevenleft{\evenleftmargin=\oddleftmargin}
\def\@setevenhead{\evenhead=\oddhead}
\def\@setevenfoot{\evenfoot=\oddfoot}

\def\pagenumstyle#1{\@setnumstyle\pagenum{#1}}

\newif\ifdraft
\def\draft{\drafttrue\leftmargin=.5in \overfullrule=5pt }

\newif\ifnumup

\def\outputstyle#1{\global\expandafter\let\expandafter
          \@outputstyle\csname#1output\endcsname
     \usename{#1setup}}

\output={\@outputstyle}

\def\normaloutput{\the\everyoutput
     \global\advance\pagenum by\@ne
     \ifodd\pagenum
          \voffset=\oddtopmargin \hoffset=\oddleftmargin
     \else\voffset=\eventopmargin \hoffset=\evenleftmargin
     \fi
     \advance\voffset by-1in  \advance\hoffset by-1in
     \count0=\pagenum
     \expandafter\shipout\pagebox
     \ifnum\outputpenalty>-\@MM\else\dosupereject\fi}

\newdimen\fullhsize
\newbox\leftpage
\newcount\leftpagenum
\newcount\outputpagenum \outputpagenum=0
\let\leftorright=L

\def\twoupoutput{\the\everyoutput
     \global\advance\pagenum by\@ne
     \if L\leftorright
          \global\setbox\leftpage=\leftline{\pagebox}%
          \global\leftpagenum=\pagenum
          \global\let\leftorright=R%
     \else\global\advance\outputpagenum by\@ne
          \ifodd\outputpagenum
               \voffset=\oddtopmargin \hoffset=\oddleftmargin
          \else\voffset=\eventopmargin \hoffset=\evenleftmargin
          \fi
          \advance\voffset by-1in  \advance\hoffset by-1in
          \count0=\leftpagenum \count1=\pagenum
          \shipout\vbox{\hbox to\fullhsize
               {\box\leftpage\hfil\leftline{\pagebox}}}%
          \global\let\leftorright=L%
     \fi
     \ifnum\outputpenalty>-\@MM
     \else\dosupereject
          \if R\leftorright
               \globaldefs=\@ne\head={\hfil}\foot={\hfil}\globaldefs=\z@
               \null\newpage
          \fi
     \fi}

\def\pagebox{\vbox{\makeheadline\pagebody\makefootline}}

\def\makeheadline{%
     \vbox to\z@{\baselinestretch=\@m
          \vskip\topskip\vskip-.708\baselineskip\vskip-\headskip
          \line{\vbox to\ht\strutbox{}%
               \ifodd\pagenum\the\oddhead\else\the\evenhead\fi}%
          \vss}%
     \nointerlineskip}

\def\pagebody{\vbox to\vsize{%
     \boxmaxdepth\maxdepth
     \ifvoid\topins\else\unvbox\topins\fi
     \depthofpage=\dp255
     \unvbox255
     \ifraggedbottom\kern-\depthofpage\vfil\fi
     \ifvoid\footins
     \else\vskip\skip\footins
          \footnoterule
          \unvbox\footins
          \vskip-\footnoteskip
     \fi}}

\def\makefootline{\baselineskip=\footskip
     \line{\ifodd\pagenum\the\oddfoot\else\the\evenfoot\fi}}


\newskip\abovechapterskip
\newskip\belowchapterskip
\newskip\abovesectionskip
\newskip\belowsectionskip
\newskip\abovesubsectionskip
\newskip\belowsubsectionskip

\def\chapterstyle#1{\global\expandafter\let\expandafter\@chapterstyle
     \csname#1text\endcsname}
\def\sectionstyle#1{\global\expandafter\let\expandafter\@sectionstyle
     \csname#1text\endcsname}
\def\subsectionstyle#1{\global\expandafter\let\expandafter\@subsectionstyle
     \csname#1text\endcsname}

\def\chapter#1{%
     \ifdim\lastskip=17sp \else\chapterbreak\vskip\abovechapterskip\fi
     \@chapterstyle{\ifblank\chapternumstyle\then
          \else\newchapternum=\next\chapternumformat\ \fi#1}%
     \nobreak\vskip\belowchapterskip\vskip17sp }

\def\section#1{%
     \ifdim\lastskip=17sp \else\sectionbreak\vskip\abovesectionskip\fi
     \@sectionstyle{\ifblank\sectionnumstyle\then
          \else\newsectionnum=\next\sectionnumformat\ \fi#1}%
     \nobreak\vskip\belowsectionskip\vskip17sp }

\def\subsection#1{%
     \ifdim\lastskip=17sp \else\subsectionbreak\vskip\abovesubsectionskip\fi
     \@subsectionstyle{\ifblank\subsectionnumstyle\then
          \else\newsubsectionnum=\next\subsectionnumformat\ \fi#1}%
     \nobreak\vskip\belowsubsectionskip\vskip17sp }


\let\TeXunderline=\underline
\let\TeXoverline=\overline
\def\underline#1{\relax\ifmmode\TeXunderline{#1}\else
     $\TeXunderline{\hbox{#1}}$\fi}
\def\overline#1{\relax\ifmmode\TeXoverline{#1}\else
     $\TeXoverline{\hbox{#1}}$\fi}

\def\baselinestretch{\afterassignment\@baselinestretch\count@}
\def\@baselinestretch{\baselineskip=\normalbaselineskip
     \divide\baselineskip by\@m\baselineskip=\count@\baselineskip
     \setbox\strutbox=\hbox{\vrule
          height.708\baselineskip depth.292\baselineskip width\z@}%
     \bigskipamount=\the\baselineskip
          plus.25\baselineskip minus.25\baselineskip
     \medskipamount=.5\baselineskip
          plus.125\baselineskip minus.125\baselineskip
     \smallskipamount=.25\baselineskip
          plus.0625\baselineskip minus.0625\baselineskip}

\def\\{\ifhmode\ifnum\lastpenalty=-\@M\else\hfil\penalty-\@M\fi\fi
     \ignorespaces}
\def\newpage{\vfil\break}

\def\lefttext#1{\par{\@text\leftskip=\z@\rightskip=\centering
     \noindent#1\par}}
\def\righttext#1{\par{\@text\leftskip=\centering\rightskip=\z@
     \noindent#1\par}}
\def\centertext#1{\par{\@text\leftskip=\centering\rightskip=\centering
     \noindent#1\par}}
\def\@text{\parindent=\z@ \parfillskip=\z@ \everypar={}%
     \spaceskip=.3333em \xspaceskip=.5em
     \def\\{\ifhmode\ifnum\lastpenalty=-\@M\else\penalty-\@M\fi\fi
          \ignorespaces}}

\def\beginleft{\par\@text\leftskip=\z@ \rightskip=\centering}
     
\def\beginright{\par\@text\leftskip=\centering\rightskip=\z@ }
     
\def\begincenter{\par\@text\leftskip=\centering\rightskip=\centering}

\def\beginnarrow{\defaultoption[\parindent]\@beginnarrow}
\def\@beginnarrow[#1]{\par\advance\leftskip by#1\advance\rightskip by#1}

\begingroup
\catcode`\[=1 \catcode`\{=11
\gdef\beginignore[\endgroup\bgroup
     \catcode`\e=0 \catcode`\\=12 \catcode`\{=11 \catcode`\f=12 \let\or=\relax
     \let\nd{ignor=\fi \let\}=\egroup
     \iffalse}
\endgroup

\long\def\marginnote#1{\leavevmode
     \edef\@marginsf{\spacefactor=\the\spacefactor\relax}%
     \ifdraft\strut\vadjust{%
          \hbox to\z@{\hskip\hsize\hskip.1in
               \vbox to\z@{\vskip-\dp\strutbox
                    \marginnoteformat
                    \vskip-\ht\strutbox
                    \noindent\strut#1\par
                    \vss}%
               \hss}}%
     \fi
     \@marginsf}


\newtoks\everybye \everybye={\par\vfil}
\outer\def\bye{\the\everybye
     \footnotecheck
     \prelabelcheck
     \streamcheck
     \supereject
     \TeXend}

\message{footnotes,}

\newcount\footnotenum \footnotenum=0
\newskip\footnoteskip
\let\@footnotelist=\empty

\def\footnotenumstyle#1{\@setnumstyle\footnotenum{#1}%
     \useafter\ifx{@footnotenumstyle}\symbols
          \global\let\@footup=\empty
     \else\global\let\@footup=\markup
     \fi}

\def\footnote{\footnotecheck\defaultoption[]\@footnote}
\def\@footnote[#1]{\@footnotemark[#1]\@footnotetext}

\def\footnotemark{\defaultoption[]\@footnotemark}
\def\@footnotemark[#1]{\let\@footsf=\empty
     \ifhmode\edef\@footsf{\spacefactor=\the\spacefactor\relax}\/\fi
     \ifnoarg#1\then
          \global\advance\footnotenum by\@ne
          \@footup{\footnotenumformat}%
          \edef\@@foota{\footnotenum=\the\footnotenum\relax}%
          \expandafter\additemR\expandafter\@footup\expandafter
               {\@@foota\footnotenumformat}\to\@footnotelist
          \global\let\@footnotelist=\@footnotelist
     \else\markup{#1}%
          \additemR\markup{#1}\to\@footnotelist
          \global\let\@footnotelist=\@footnotelist
     \fi
     \@footsf}

\def\footnotetext{%
     \ifx\@footnotelist\empty\err@extrafootnotetext\else\@footnotetext\fi}
\def\@footnotetext{%
     \getitemL\@footnotelist\to\@@foota
     \global\let\@footnotelist=\@footnotelist
     \insert\footins\bgroup
     \footnoteformat
     \splittopskip=\ht\strutbox\splitmaxdepth=\dp\strutbox
     \interlinepenalty=\interfootnotelinepenalty\floatingpenalty=\@MM
     \noindent\llap{\@@foota}\strut
     \bgroup\aftergroup\@footnoteend
     \let\@@scratcha=}
\def\@footnoteend{\strut\par\vskip\footnoteskip\egroup}

\def\footnoterule{\normalfonts
     \kern-.3em \hrule width2in height.04em \kern .26em }

\def\footnotecheck{%
     \ifx\@footnotelist\empty
     \else\err@extrafootnotemark
          \global\let\@footnotelist=\empty
     \fi}

\message{labels,}

\let\@@labeldef=\xdef
\newif\if@labelfile
\newwrite\@labelfile
\let\@prelabellist=\empty

\def\label#1#2{\trim#1\to\@@labarg\edef\@@labtext{#2}%
     \edef\@@labname{lab@\@@labarg}%
     \useafter\ifundefined\@@labname\then\else\@yeslab\fi
     \useafter\@@labeldef\@@labname{#2}%
     \ifstreaming
          \expandafter\toks@\expandafter\expandafter\expandafter
               {\csname\@@labname\endcsname}%
          \immediate\write\streamout{\noexpand\label{\@@labarg}{\the\toks@}}%
     \fi}
\def\@yeslab{%
     \useafter\ifundefined{if\@@labname}\then
          \err@labelredef\@@labarg
     \else\useif{if\@@labname}\then
               \err@labelredef\@@labarg
          \else\global\usename{\@@labname true}%
               \useafter\ifundefined{pre\@@labname}\then
               \else\useafter\ifx{pre\@@labname}\@@labtext
                    \else\err@badlabelmatch\@@labarg
                    \fi
               \fi
               \if@labelfile
               \else\global\@labelfiletrue
                    \immediate\write\sixt@@n{--Creating file \jobname.lab}%
                    \immediate\openout\@labelfile=\jobname.lab
               \fi
               \immediate\write\@labelfile
                    {\noexpand\prelabel{\@@labarg}{\@@labtext}}%
          \fi
     \fi}

\def\putlab#1{\trim#1\to\@@labarg\edef\@@labname{lab@\@@labarg}%
     \useafter\ifundefined\@@labname\then\@nolab\else\usename\@@labname\fi}
\def\@nolab{%
     \useafter\ifundefined{pre\@@labname}\then
          \undefinedlabelformat
          \err@needlabel\@@labarg
          \useafter\xdef\@@labname{\undefinedlabelformat}%
     \else\usename{pre\@@labname}%
          \useafter\xdef\@@labname{\usename{pre\@@labname}}%
     \fi
     \useafter\newif{if\@@labname}%
     \expandafter\additemR\@@labarg\to\@prelabellist}

\def\prelabel#1{\useafter\gdef{prelab@#1}}

\def\ifundefinedlabel#1\then{%
     \expandafter\ifx\csname lab@#1\endcsname\relax}
\def\useiflab#1\then{\csname iflab@#1\endcsname}

\def\prelabelcheck{{%
     \def\^^\##1{\useiflab{##1}\then\else\err@undefinedlabel{##1}\fi}%
     \@prelabellist}}

\message{equation numbering,}

\newcount\chapternum
\newcount\sectionnum
\newcount\subsectionnum
\newcount\equationnum
\newcount\subequationnum
\newcount\figurenum
\newcount\subfigurenum
\newcount\tablenum
\newcount\subtablenum

\newif\if@subeqncount
\newif\if@subfigcount
\newif\if@subtblcount

\def\newchapternum{\newsectionnum=\z@\@resetnum\chapternum}
\def\newsectionnum{\newsubsectionnum=\z@\@resetnum\sectionnum}
\def\newsubsectionnum{\newequationnum=\z@\newfigurenum=\z@\newtablenum=\z@
     \@resetnum\subsectionnum}
\def\newequationnum{\newsubequationnum=\z@\@resetnum\equationnum}
\def\newsubequationnum{\@resetnum\subequationnum}
\def\newfigurenum{\newsubfigurenum=\z@\@resetnum\figurenum}
\def\newsubfigurenum{\@resetnum\subfigurenum}
\def\newtablenum{\newsubtablenum=\z@\@resetnum\tablenum}
\def\newsubtablenum{\@resetnum\subtablenum}

\def\@resetnum#1{\global\advance#1by1 \edef\next{\the#1\relax}\global#1}

\newchapternum=0

\def\chapternumstyle#1{\@setnumstyle\chapternum{#1}}
\def\sectionnumstyle#1{\@setnumstyle\sectionnum{#1}}
\def\subsectionnumstyle#1{\@setnumstyle\subsectionnum{#1}}
\def\equationnumstyle#1{\@setnumstyle\equationnum{#1}}
\def\subequationnumstyle#1{\@setnumstyle\subequationnum{#1}%
     \ifblank\subequationnumstyle\then\global\@subeqncountfalse\fi
     \ignorespaces}
\def\figurenumstyle#1{\@setnumstyle\figurenum{#1}}
\def\subfigurenumstyle#1{\@setnumstyle\subfigurenum{#1}%
     \ifblank\subfigurenumstyle\then\global\@subfigcountfalse\fi
     \ignorespaces}
\def\tablenumstyle#1{\@setnumstyle\tablenum{#1}}
\def\subtablenumstyle#1{\@setnumstyle\subtablenum{#1}%
     \ifblank\subtablenumstyle\then\global\@subtblcountfalse\fi
     \ignorespaces}

\def\eqnlabel#1{%
     \if@subeqncount
          \newsubequationnum=\next
     \else\newequationnum=\next
          \ifblank\subequationnumstyle\then
          \else\global\@subeqncounttrue
               \newsubequationnum=\@ne
          \fi
     \fi
     \label{#1}{\puteqnformat}(\puteqn{#1})%
     \ifdraft\rlap{\hskip.1in{\tt#1}}\fi}

\let\puteqn=\putlab

\def\equation#1#2{\useafter\gdef{eqn@#1}{#2\eqno\eqnlabel{#1}}}
\def\Equation#1{\useafter\gdef{eqn@#1}}

\def\putequation#1{\useafter\ifundefined{eqn@#1}\then
     \err@undefinedeqn{#1}\else\usename{eqn@#1}\fi}

\def\eqnseriesstyle#1{\gdef\@eqnseriesstyle{#1}}
\def\begineqnseries{\subequationnumstyle{\@eqnseriesstyle}%
     \defaultoption[]\@begineqnseries}
\def\@begineqnseries[#1]{\edef\@@eqnname{#1}}
\def\endeqnseries{\subequationnumstyle{blank}%
     \expandafter\ifnoarg\@@eqnname\then
     \else\label\@@eqnname{\puteqnformat}%
     \fi
     \aftergroup\ignorespaces}

\def\figlabel#1{%
     \if@subfigcount
          \newsubfigurenum=\next
     \else\newfigurenum=\next
          \ifblank\subfigurenumstyle\then
          \else\global\@subfigcounttrue
               \newsubfigurenum=\@ne
          \fi
     \fi
     \label{#1}{\putfigformat}\putfig{#1}%
     {\def\marginnoteformat{\tt}\marginnote{#1}}}

\let\putfig=\putlab

\def\figseriesstyle#1{\gdef\@figseriesstyle{#1}}
\def\beginfigseries{\subfigurenumstyle{\@figseriesstyle}%
     \defaultoption[]\@beginfigseries}
\def\@beginfigseries[#1]{\edef\@@figname{#1}}
\def\endfigseries{\subfigurenumstyle{blank}%
     \expandafter\ifnoarg\@@figname\then
     \else\label\@@figname{\putfigformat}%
     \fi
     \aftergroup\ignorespaces}

\def\tbllabel#1{%
     \if@subtblcount
          \newsubtablenum=\next
     \else\newtablenum=\next
          \ifblank\subtablenumstyle\then
          \else\global\@subtblcounttrue
               \newsubtablenum=\@ne
          \fi
     \fi
     \label{#1}{\puttblformat}\puttbl{#1}%
     {\def\marginnoteformat{\tt}\marginnote{#1}}}

\let\puttbl=\putlab

\def\tblseriesstyle#1{\gdef\@tblseriesstyle{#1}}
\def\begintblseries{\subtablenumstyle{\@tblseriesstyle}%
     \defaultoption[]\@begintblseries}
\def\@begintblseries[#1]{\edef\@@tblname{#1}}
\def\endtblseries{\subtablenumstyle{blank}%
     \expandafter\ifnoarg\@@tblname\then
     \else\label\@@tblname{\puttblformat}%
     \fi
     \aftergroup\ignorespaces}

\message{reference numbering,}

\newcount\referencenum \referencenum=0
\newcount\@@prerefcount \@@prerefcount=0
\newcount\@@thisref
\newcount\@@lastref
\newcount\@@loopref
\newcount\@@refseq
\newdimen\refnumindent
\let\@undefreflist=\empty

\def\referencenumstyle#1{\@setnumstyle\referencenum{#1}}

\def\referencestyle#1{\usename{@ref#1}}

\def\@refsequential{%
     \gdef\@refpredef##1{\global\advance\referencenum by\@ne
          \let\^^\=0\label{##1}{\^^\{\the\referencenum}}%
          \useafter\gdef{ref@\the\referencenum}{{##1}{\undefinedlabelformat}}}%
     \gdef\@reference##1##2{%
          \ifundefinedlabel##1\then
          \else\def\^^\####1{\global\@@thisref=####1\relax}\putlab{##1}%
               \useafter\gdef{ref@\the\@@thisref}{{##1}{##2}}%
          \fi}%
     \gdef\endputreferences{%
          \loop\ifnum\@@loopref<\referencenum
                    \advance\@@loopref by\@ne
                    \expandafter\expandafter\expandafter\@printreference
                         \csname ref@\the\@@loopref\endcsname
          \repeat
          \par}}

\def\@refpreordered{%
     \gdef\@refpredef##1{\global\advance\referencenum by\@ne
          \additemR##1\to\@undefreflist}%
     \gdef\@reference##1##2{%
          \ifundefinedlabel##1\then
          \else\global\advance\@@loopref by\@ne
               {\let\^^\=0\label{##1}{\^^\{\the\@@loopref}}}%
               \@printreference{##1}{##2}%
          \fi}
     \gdef\endputreferences{%
          \def\^^\####1{\useiflab{####1}\then
               \else\reference{####1}{\undefinedlabelformat}\fi}%
          \@undefreflist
          \par}}

\def\beginprereferences{\par
     \def\reference##1##2{\global\advance\referencenum by1\@ne
          \let\^^\=0\label{##1}{\^^\{\the\referencenum}}%
          \useafter\gdef{ref@\the\referencenum}{{##1}{##2}}}}
\def\endprereferences{\global\@@prerefcount=\the\referencenum\par}

\def\beginputreferences{\par
     \refnumindent=\z@\@@loopref=\z@
     \loop\ifnum\@@loopref<\referencenum
               \advance\@@loopref by\@ne
               \setbox\z@=\hbox{\referencenum=\@@loopref
                    \referencenumformat\enskip}%
               \ifdim\wd\z@>\refnumindent\refnumindent=\wd\z@\fi
     \repeat
     \putreferenceformat
     \@@loopref=\z@
     \loop\ifnum\@@loopref<\@@prerefcount
               \advance\@@loopref by\@ne
               \expandafter\expandafter\expandafter\@printreference
                    \csname ref@\the\@@loopref\endcsname
     \repeat
     \let\reference=\@reference}

\def\@printreference#1#2{\ifx#2\undefinedlabelformat\err@undefinedref{#1}\fi
     \noindent\ifdraft\rlap{\hskip\hsize\hskip.1in \tt#1}\fi
     \llap{\referencenum=\@@loopref\referencenumformat\enskip}#2\par}

\def\reference#1#2{{\par\refnumindent=\z@\putreferenceformat\noindent#2\par}}

\def\putref#1{\trim#1\to\@@refarg
     \expandafter\ifnoarg\@@refarg\then
          \toks@={\relax}%
     \else\@@lastref=-\@m\def\@@refsep{}\def\@more{\@nextref}%
          \toks@={\@nextref#1,,}%
     \fi\the\toks@}
\def\@nextref#1,{\trim#1\to\@@refarg
     \expandafter\ifnoarg\@@refarg\then
          \let\@more=\relax
     \else\ifundefinedlabel\@@refarg\then
               \expandafter\@refpredef\expandafter{\@@refarg}%
          \fi
          \def\^^\##1{\global\@@thisref=##1\relax}%
          \global\@@thisref=\m@ne
          \setbox\z@=\hbox{\putlab\@@refarg}%
     \fi
     \advance\@@lastref by\@ne
     \ifnum\@@lastref=\@@thisref\advance\@@refseq by\@ne\else\@@refseq=\@ne\fi
     \ifnum\@@lastref<\z@
     \else\ifnum\@@refseq<\thr@@
               \@@refsep\def\@@refsep{,}%
               \ifnum\@@lastref>\z@
                    \advance\@@lastref by\m@ne
                    {\referencenum=\@@lastref\putrefformat}%
               \else\undefinedlabelformat
               \fi
          \else\def\@@refsep{--}%
          \fi
     \fi
     \@@lastref=\@@thisref
     \@more}

\message{streaming,}

\newif\ifstreaming

\def\streamto{\defaultoption[\jobname]\@streamto}
\def\@streamto[#1]{\global\streamingtrue
     \immediate\write\sixt@@n{--Streaming to #1.str}%
     \newwrite\streamout\immediate\openout\streamout=#1.str }

\def\streamfrom{\defaultoption[\jobname]\@streamfrom}
\def\@streamfrom[#1]{\newread\streamin\openin\streamin=#1.str
     \ifeof\streamin
          \expandafter\err@nostream\expandafter{#1.str}%
     \else\immediate\write\sixt@@n{--Streaming from #1.str}%
          \let\@@labeldef=\gdef
          \ifstreaming
               \edef\@elc{\endlinechar=\the\endlinechar}%
               \endlinechar=\m@ne
               \loop\read\streamin to\@@scratcha
                    \ifeof\streamin
                         \streamingfalse
                    \else\toks@=\expandafter{\@@scratcha}%
                         \immediate\write\streamout{\the\toks@}%
                    \fi
                    \ifstreaming
               \repeat
               \@elc
               \input #1.str
               \streamingtrue
          \else\input #1.str
          \fi
          \let\@@labeldef=\xdef
     \fi}

\def\streamcheck{\ifstreaming
     \immediate\write\streamout{\pagenum=\the\pagenum}%
     \immediate\write\streamout{\footnotenum=\the\footnotenum}%
     \immediate\write\streamout{\referencenum=\the\referencenum}%
     \immediate\write\streamout{\chapternum=\the\chapternum}%
     \immediate\write\streamout{\sectionnum=\the\sectionnum}%
     \immediate\write\streamout{\subsectionnum=\the\subsectionnum}%
     \immediate\write\streamout{\equationnum=\the\equationnum}%
     \immediate\write\streamout{\subequationnum=\the\subequationnum}%
     \immediate\write\streamout{\figurenum=\the\figurenum}%
     \immediate\write\streamout{\subfigurenum=\the\subfigurenum}%
     \immediate\write\streamout{\tablenum=\the\tablenum}%
     \immediate\write\streamout{\subtablenum=\the\subtablenum}%
     \immediate\closeout\streamout
     \fi}


\def\err@badtypesize{%
     \errhelp={The limited availability of certain fonts requires^^J%
          that the base type size be 10pt, 12pt, or 14pt.^^J}%
     \errmessage{--Illegal base type size}}

\def\err@badsizechange{\immediate\write\sixt@@n
     {--Size change not allowed in math mode, ignored}}

\def\err@sizetoolarge#1{\immediate\write\sixt@@n
     {--\noexpand#1 too big, substituting HUGE}}

\def\err@sizenotavailable#1{\immediate\write\sixt@@n
     {--Size not available, \noexpand#1 ignored}}

\def\err@fontnotavailable#1{\immediate\write\sixt@@n
     {--Font not available, \noexpand#1 ignored}}

\def\err@sltoit{\immediate\write\sixt@@n
     {--Style \noexpand\sl not available, substituting \noexpand\it}%
     \it}

\def\err@bfstobf{\immediate\write\sixt@@n
     {--Style \noexpand\bfs not available, substituting \noexpand\bf}%
     \bf}

\def\err@badgroup#1#2{%
     \errhelp={The block you have just tried to close was not the one^^J%
          most recently opened.^^J}%
     \errmessage{--\noexpand\end{#1} doesn't match \noexpand\begin{#2}}}

\def\err@badcountervalue#1{\immediate\write\sixt@@n
     {--Counter (#1) out of bounds}}

\def\err@extrafootnotemark{\immediate\write\sixt@@n
     {--\noexpand\footnotemark command
          has no corresponding \noexpand\footnotetext}}

\def\err@extrafootnotetext{%
     \errhelp{You have given a \noexpand\footnotetext command without first
          specifying^^Ja \noexpand\footnotemark.^^J}%
     \errmessage{--\noexpand\footnotetext command has no corresponding
          \noexpand\footnotemark}}

\def\err@labelredef#1{\immediate\write\sixt@@n
     {--Label "#1" redefined}}

\def\err@badlabelmatch#1{\immediate\write\sixt@@n
     {--Definition of label "#1" doesn't match value in \jobname.lab}}

\def\err@needlabel#1{\immediate\write\sixt@@n
     {--Label "#1" cited before its definition}}

\def\err@undefinedlabel#1{\immediate\write\sixt@@n
     {--Label "#1" cited but never defined}}

\def\err@undefinedeqn#1{\immediate\write\sixt@@n
     {--Equation "#1" not defined}}

\def\err@undefinedref#1{\immediate\write\sixt@@n
     {--Reference "#1" not defined}}

\def\err@nostream#1{%
     \errhelp={You have tried to input a stream file that doesn't exist.^^J}%
     \errmessage{--Stream file #1 not found}}

\message{jyTeX initialization}

\everyjob{\immediate\write16{--jyTeX version \fmtversion}%
     \edef\@@jobname{\jobname}%
     \edef\jobname{\@@jobname}%
     \settime
     \openin0=\jobname.lab
     \ifeof0
     \else\closein0
          \immediate\write16{--Getting labels from file \jobname.lab}%
          \input\jobname.lab
     \fi}


\def\fixedskipslist{%
     \^^\{\topskip}%
     \^^\{\splittopskip}%
     \^^\{\maxdepth}%
     \^^\{\skip\topins}%
     \^^\{\skip\footins}%
     \^^\{\headskip}%
     \^^\{\footskip}}

\def\scalingskipslist{%
     \^^\{\p@renwd}%
     \^^\{\delimitershortfall}%
     \^^\{\nulldelimiterspace}%
     \^^\{\scriptspace}%
     \^^\{\jot}%
     \^^\{\normalbaselineskip}%
     \^^\{\normallineskip}%
     \^^\{\normallineskiplimit}%
     \^^\{\baselineskip}%
     \^^\{\lineskip}%
     \^^\{\lineskiplimit}%
     \^^\{\bigskipamount}%
     \^^\{\medskipamount}%
     \^^\{\smallskipamount}%
     \^^\{\parskip}%
     \^^\{\parindent}%
     \^^\{\abovedisplayskip}%
     \^^\{\belowdisplayskip}%
     \^^\{\abovedisplayshortskip}%
     \^^\{\belowdisplayshortskip}%
     \^^\{\abovechapterskip}%
     \^^\{\belowchapterskip}%
     \^^\{\abovesectionskip}%
     \^^\{\belowsectionskip}%
     \^^\{\abovesubsectionskip}%
     \^^\{\belowsubsectionskip}}


\def\twoupsetup{
     \topmargin=.75in
     \leftmargin=.5in
     \vsize=6.9in
     \hsize=4.75in
     \fullhsize=10in
     \let\draft=\relax}

\outputstyle{normal}                             

\def\marginnoteformat{\subscriptsize             
     \hsize=1in \baselinestretch=1000 \everypar={}%
     \tolerance=5000 \hbadness=5000 \parskip=0pt \parindent=0pt
     \leftskip=0pt \rightskip=0pt \raggedright}


\head={\ifnumup\normalfonts\hfill\numstyle\pagenum  
        \else\hfil\fi}

\foot={\ifnumup\hfil\else
\hfil\normalfonts\numstyle\pagenum\hfil\fi}  

\normalbaselineskip=12pt                         
\normallineskip=0pt                              
\normallineskiplimit=0pt                         
\normalbaselines                                 

\topskip=.85\baselineskip
\splittopskip=\topskip
\headskip=2\baselineskip
\footskip=\headskip

\pagenumstyle{arabic}                            

\parskip=0pt                                     
\parindent=20pt                                  

\baselinestretch=1000                            


\chapterstyle{left}                              
\chapternumstyle{blank}                          
\def\chapterbreak{\newpage}                      
\abovechapterskip=0pt                            
\belowchapterskip=1.5\baselineskip               
     plus.38\baselineskip minus.38\baselineskip
\def\chapternumformat{\numstyle\chapternum.}     

\sectionstyle{left}                              
\sectionnumstyle{blank}                          
\def\sectionbreak{\vskip0pt plus4\baselineskip\penalty-100
     \vskip0pt plus-4\baselineskip}              
\abovesectionskip=1.5\baselineskip               
     plus.38\baselineskip minus.38\baselineskip
\belowsectionskip=\the\baselineskip              
     plus.25\baselineskip minus.25\baselineskip
\def\sectionnumformat{
     \ifblank\chapternumstyle\then\else\numstyle\chapternum.\fi
     \numstyle\sectionnum.}

\subsectionstyle{left}                           
\subsectionnumstyle{blank}                       
\def\subsectionbreak{\vskip0pt plus4\baselineskip\penalty-100
     \vskip0pt plus-4\baselineskip}              
\abovesubsectionskip=\the\baselineskip           
     plus.25\baselineskip minus.25\baselineskip
\belowsubsectionskip=.75\baselineskip            
     plus.19\baselineskip minus.19\baselineskip
\def\subsectionnumformat{
     \ifblank\chapternumstyle\then\else\numstyle\chapternum.\fi
     \ifblank\sectionnumstyle\then\else\numstyle\sectionnum.\fi
     \numstyle\subsectionnum.}


\footnotenumstyle{symbols}                       
\footnoteskip=0pt                                
\def\footnotenumformat{\numstyle\footnotenum}    
\def\footnoteformat{\footnotesize                
     \everypar={}\parskip=0pt \parfillskip=0pt plus1fil
     \leftskip=1em \rightskip=0pt
     \spaceskip=0pt \xspaceskip=0pt
     \def\\{\ifhmode\ifnum\lastpenalty=-10000
          \else\hfil\penalty-10000 \fi\fi\ignorespaces}}


\def\undefinedlabelformat{$\bullet$}             


\equationnumstyle{arabic}                        
\subequationnumstyle{blank}                      
\figurenumstyle{arabic}                          
\subfigurenumstyle{blank}                        
\tablenumstyle{arabic}                           
\subtablenumstyle{blank}                         

\eqnseriesstyle{alphabetic}                      
\figseriesstyle{alphabetic}                      
\tblseriesstyle{alphabetic}                      

\def\puteqnformat{\hbox{
     \ifblank\chapternumstyle\then\else\numstyle\chapternum.\fi
     \ifblank\sectionnumstyle\then\else\numstyle\sectionnum.\fi
     \ifblank\subsectionnumstyle\then\else\numstyle\subsectionnum.\fi
     \numstyle\equationnum
     \numstyle\subequationnum}}
\def\putfigformat{\hbox{
     \ifblank\chapternumstyle\then\else\numstyle\chapternum.\fi
     \ifblank\sectionnumstyle\then\else\numstyle\sectionnum.\fi
     \ifblank\subsectionnumstyle\then\else\numstyle\subsectionnum.\fi
     \numstyle\figurenum
     \numstyle\subfigurenum}}
\def\puttblformat{\hbox{
     \ifblank\chapternumstyle\then\else\numstyle\chapternum.\fi
     \ifblank\sectionnumstyle\then\else\numstyle\sectionnum.\fi
     \ifblank\subsectionnumstyle\then\else\numstyle\subsectionnum.\fi
     \numstyle\tablenum
     \numstyle\subtablenum}}


\referencestyle{sequential}                      
\referencenumstyle{arabic}                       
\def\putrefformat{\numstyle\referencenum}        
\def\referencenumformat{\numstyle\referencenum.} 
\def\putreferenceformat{
     \everypar={\hangindent=1em \hangafter=1 }%
     \def\\{\hfil\break\null\hskip-1em \ignorespaces}%
     \leftskip=\refnumindent\parindent=0pt \interlinepenalty=1000 }


\normalsize


\def\fmtversion{2.6M (June 1992)}
%
%
%

\def\PRL#1#2#3{{\sl Phys. Rev. Lett.} {\bf#1} (#2) #3}
\def\PR#1#2#3{{\sl Phys. Rev.} {\bf#1} (#2) #3}

\def\CMP#1#2#3{{\sl Comm. Math. Phys.} {\bf #1} (#2) #3}

\def\PLA#1#2#3{{\sl Phys. Lett.} {\bf #1A} (#2) #3}

\def\IJMPB#1#2#3{{\sl Int. J. Mod. Phys.} {\bf B#1} (#2) #3}

\def\JPA#1#2#3{{\sl J. Physics} {\bf A#1} (#2) #3}

\def\Phy#1#2#3{{\sl Physica} {\bf #1} (#2) #3}

\def\SPD#1#2#3{{\sl Sov. Phys. Dokl.} {\bf #1} (#2) #3}

\def\upref#1/{\markup{[\putref{#1}]}}
\def\Idoubled#1{{\rm I\kern-.22em #1}}
\def\Odoubled#1{{\setbox0=\hbox{\rm#1}%
     \dimen@=\ht0 \dimen@ii=.04em \advance\dimen@ by-\dimen@ii
     \rlap{\kern.26em \vrule height\dimen@ depth-\dimen@ii width.075em}\box0}}

\def\Complex{\Odoubled C}
\catcode`\@=12

\typesize=12pt
\draft
\equation{R}{R(\lambda ,\mu) =\left(\matrix{f(\mu ,\lambda) &0&0&0\cr
0&g(\mu ,\lambda)&1&0\cr 0&1&g(\mu ,\lambda)&0\cr 0&0&0&f(\mu ,\lambda)\cr
}\right) ,}

\equation{fgxxz}{f(\lambda ,\mu) = {\sinh(\lambda -\mu +2i\eta)\over
\sinh(\lambda -\mu)} \ ,\qquad g(\lambda ,\mu) = {i\sin(2\eta)\over
\sinh(\lambda -\mu)} \ ,}

\equation{fgxxx}{f(\lambda ,\mu) = 1+{i\over \lambda -\mu} \ ,\qquad
g(\lambda ,\mu) = {i\over \lambda -\mu} \ .}

\equation{lop}{L^{xxz}_n(\mu) =\left(\matrix{\sinh(\mu -i\eta\sigma_n^z)
&-i\sin(2\eta) \sigma_n^-\cr -i\sin(2\eta) \sigma_n^+
&\sinh(\mu +i\eta\sigma_n^z)\cr}\right)\ , \quad
L^{xxx}_n(\mu) =\mu -{i\over 2} \left(\matrix{\sigma_n^z
&2\sigma_n^-\cr 2\sigma_n^+ &-\sigma_n^z\cr}\right)\ .}

\equation{intL}{R(\lambda -\mu) \left(L_n(\lambda)\otimes L_n(\mu)\right)
=  \left(L_n(\mu)\otimes L_n(\lambda)\right) R(\lambda -\mu) \ .}

\equation{intL2}{R(\lambda -\mu) \left(L_n(\lambda -\nu_n)\otimes
L_n(\mu -\nu_n)\right) =  \left(L_n(\mu -\nu_n)\otimes L_n(\lambda
-\nu_n)\right) R(\lambda -\mu) \ .}

\equation{monodr}{T(\mu) = L_L(\mu)L_{L-1}(\mu)\ldots L_1(\mu) =
\left(\matrix{A(\mu)&B(\mu)\cr C(\mu)&D(\mu)\cr}\right) \ .}

\equation{Tinhom}{T_{inh}(\l) = L_L(\l -\nu_L)L_{L-1}(\l
-\nu_{L-1})\ldots L_1(\l -\nu_1) = \left(\matrix{\A{\mu}&\B{\mu}\cr
\C{\mu}&\D{\mu}\cr}\right) \ .}

\equation{intT}{R(\lambda -\mu) \left(T(\lambda)\otimes T(\mu)\right)
=  \left(T(\mu)\otimes T(\lambda)\right) R(\lambda -\mu) \ .}

\equation{intTme}{\eqalign{[B(\l),B(\m)]&=0=[C(\l),C(\m)]\cr
[B(\l),C(\m)]&= g(\l ,\m) \left(D(\l) A(\m) - D(\m) A(\l) \right)\cr
D(\m)B(\l)&= f(\l ,\m) B(\l)D(\m) + g(\m ,\l)B(\m)D(\l) \cr
A(\m)B(\l)&= f(\m ,\l) B(\l)A(\m) + g(\l ,\m)B(\m)A(\l) .\cr}}

\equation{Hxxz}{H_{xxz} = -2i\sin(2\eta) {\partial\over \partial\mu}
\ln(\tau_{xxz}(\mu))\bigg|_{\mu=-i\eta} - 2L\cos(2\eta) -2hS^z\ .}

\equation{ADxxz}{\eqalign{A(\mu)\vac &= a(\mu)\vac\ ,\quad a(\mu) =
\left(\sinh(\mu -i\eta)\right)^L\ ,\cr
D(\mu)\vac &= d(\mu)\vac\ ,\quad d(\mu) =
\left(\sinh(\mu +i\eta)\right)^L,\ \cr
C(\mu)\vac &=0\ ,\cr
B(\mu)\vac &\neq0\ ,\cr}}

\equation{adxxx}{a(\mu)= \left(\mu -{i\over 2}\right)^L\ ,\quad
d(\mu)= \left(\mu +{i\over 2}\right)^L\ .}

\equation{H}{H=\sum_{j=1}^L
\sigma^x_j\sigma^x_{j+1}+\sigma^y_j\sigma^y_{j+1} +\Delta \left(
\sigma^z_j\sigma^z_{j+1}-1\right) - h\sum_{j=1}^L \sigma_j^z\ ,}

\equation{g}{G(m)=4\langle q_mq_1\rangle-4\langle q_1\rangle +1\ ,}

\equation{szsz}{G(m) = 2{\widehat\Delta}\langle
(Q_1(m))^2\rangle + 1-4\int_{-\Lambda}^\Lambda d\lambda\
\rho(\lambda)\ .}

\equation{genfu}{F(\alpha,m):=\langle\exp(\alpha Q_1(m))\rangle:=
\frac{\langle 0|\prod_{j=1}^{{N}}C(\lambda_j)
\exp(\alpha\sum_{l=1}^m\sigma_l^-\sigma_l^+)\prod_{k=1}^{{N}}
B(\lambda_k)|0\rangle}{\langle
0|\prod_{j=1}^{{N}}C(\lambda_j)\prod_{k=1}^{{N}}B(\lambda_k)|0\rangle}\ ,}

\equation{t12}{T(\mu) = T(2,\mu)T(1,\mu)\ ,\
T(i,\mu) =\left(\matrix{A_i(\mu) &B_i(\mu)\cr C_i(\mu)
&D_i(\mu)\cr}\right)\ (i=1,2)\ .}

\equation{t12l}{\eqalign{T(2,\mu) &= L_L(\mu)L_{L-1}(\mu)\ldots
L_n(\mu)\ ,\cr T(1,\mu) &= L_{n-1}(\mu)L_{n-2}(\mu)\ldots L_1(\mu)\
.\cr}}

\equation{abcdi}{\eqalign{A_i(\mu)\vac_i &= a_i(\mu)\vac\ ,\quad
D_i(\mu)\vac_i = d_i(\mu)\vac_i\ ,\cr
C(\mu)\vac_i &=0\ ,\quad B_i(\mu)\vac_i\neq 0\ ,\cr}}

\equation{bbb}{\eqalign{\prod_{j=1}^N B(\lambda_j)\vac &=
\sum_{I,II}\prod_{j\in I}^{n_1}\prod_{k\in II}^{n_2} a_2(\lambda_j^I)
d_1(\lambda_k^{II})\cr &\quad \times f(\lambda_j^I,\lambda_k^{II})
\left(B_2(\lambda_k^{II})\vac_2\right)\otimes
\left(B_1(\lambda_j^{I})\vac_1\right)\ ,\cr}}

\equation{ccc}{\eqalign{\langle 0|\prod_{j=1}^N C(\lambda_j) &=
\sum_{I,II}\prod_{j\in I}^{n_1}\prod_{k\in II}^{n_2} d_2(\lambda_j^I)
a_1(\lambda_k^{II})\cr &\quad \times f(\lambda_k^{II},\lambda_j^{I})
\left({_1}\dvac C_1(\lambda_j^{I})\right)\otimes
\left({_2}\dvac C_2(\lambda_k^{II})\right)\ .\cr}}

\equation{states}{\Psi_N(\lambda_1 ,\ldots ,\lambda_N) = \prod_{j=1}^N
B(\lambda_j)\vac\ .}

\equation{bae}{{a(\lambda_j)\over d(\lambda_j)} =
\prod_{\scriptstyle k=1\atop \scriptstyle k\neq j}^N
{f(\lambda_k,\lambda_j)\over f(\lambda_j,\lambda_k)}\ , j=1,\ldots ,N\ .}

\equation{gsie}{2\pi \rho(\lambda) - \int_{-\Lambda}^{\Lambda} d\mu\
K(\lambda ,\mu)\ \rho(\mu)\ = D(\l)\ ,}

\equation{kernel}{\eqalign{K (\m, \l ) &= \frac{\sin(4\eta)}{\sinh(\mu
- \lambda +2i\eta)\sinh(\mu - \lambda -2i\eta)}\ ,\ D(\l) =
{-\sin(2\eta)\over \sinh(\lambda -i\eta) \sinh(\lambda +i\eta) }\ .\cr
}}

\equation{kernelXXX}{\eqalign{K_{XXX}(\m, \l ) &= -\frac{2}{(\mu -
\lambda)^2+1}
\ ,\qquad D_{XXX}(\l) = {1\over\l^2 + \frac{1}{4}}\ .\cr}}

\equation{gsie2}{\eps(\lambda) - {1\over
2\pi}\int_{-\Lambda}^{\Lambda} d\mu\ K(\lambda ,\mu)\ \eps(\mu)\ =
2h -{2 (\sin(2\eta))^2\over \sinh(\lambda -i\eta) \sinh(\lambda
+i\eta) }\ .}

\equation{genfu2}{\eqalign{F(\alpha ,m) &= {1\over \sigma_{N}}\sum_{}
{_1}\dvac\prod_{I_C} C_1(\l^C_{I_C}) \prod_{{I_B}} B_1(\l^B_{I_B})\vac_1\
{_2}\dvac\prod_{II_C} C_2(\l^C_{II_C}) \prod_{II_B} B_2(\l^B_{II_B})\vac_2\
e^{\alpha n_1}\cr
&\times \prod_{I_B,I_C} a_2(\l^B_{I_B})d_2(\l^C_{I_C})\prod_{II_B,II_C}
a_1(\l^C_{II_C})d_1(\l^B_{II_B}) \prod_{I_B,II_B} f(\l^B_{I_B} ,\l^B_{II_B})
\prod_{I_C,II_C} f(\l^C_{II_C} ,\l^C_{I_C}) \ ,\cr}}

\equation{Sn}{\sigma_{N} = {\langle 0| \prod_{j=1}^{{N}} C(\lambda_j)
\prod_{k=1}^{{N}}B(\lambda_k)|0\rangle}\ .}

\equation{detZ1}{\eqalign{
{\bar{\dvac}} \prod_{j=1}^N \B{\lambda_j}{\vac} &=
{\dvac} \prod_{j=1}^N \C{\lambda_j}{\bar{\vac}} \cr
&= (-1)^N\prod_{\alpha =1}^N \prod_{k =1}^N \sinh(\lambda_\alpha
-\nu_k-i\eta)\ \sinh(\lambda_\alpha -\nu_k+i\eta) \cr
&\quad \times\left(\prod_{1\leq\alpha <\beta\leq N} \sinh(\l_\alpha
-\l_\beta) \prod_{1\leq k <l\leq N} \sinh(\nu_l -\nu_k) \right)^{-1}{\rm
det}({\cal M}) \ ,\cr}}

\equation{detZ2}{{\cal M}_{\alpha k} = {i\sin(2\eta)\over
\sinh(\l_\alpha -\nu_k -i\eta)\sinh(\l_\alpha -\nu_k +i\eta)}\ .}

\equation{Ks}{{S}_N = \sum_{A,D} \prod_{j=1}^N a(\l_j^A)\prod_{k=1}^N
d(\l_k^D) K_N\left(\matrix{
\{\l^C\}&\{\l^B\}\cr\{\l^A\}&\{\l^D\}\cr}\right)\ ,}

\equation{scpr}{{S}_N = {\langle 0| \prod_{j=1}^{N} C(\lambda_j^C)
\prod_{k=1}^{N}B(\lambda_k^B)|0\rangle}\ .}

\equation{Kcbcb}{\eqalign{K_N\left(\matrix{
\{\l^C\}&\{\l^B\}\cr\{\l^C\}&\{\l^B\}\cr}\right) &= \left(\prod_{j>k}
g(\l_j^B, \l_k^B)g(\l_k^C, \l_j^C)\right) \prod_{j,k} h(\l_j^C,
\l_k^B) {\rm det}(M^{B}_{C})\, \cr
h(\mu,\nu) &= {f(\mu,\nu)\over g(\mu,\nu)}\ ,\quad \left(M^B_C\right)_{jk}
= {g(\l_j^C,\l^B_k)\over h(\l_j^C,\l^B_k)} = t(\l_j^C,\l^B_k)\ .\cr}}

\equation{Kcbcbxxz}{\eqalign{
h(\l,\m) &= {\sinh(\l-\m+2i\eta)\over i\sin(2\eta)}\ ,\
t(\l,\m)= {(i\sin(2\eta))^2\over
\sinh(\l-\m+2i\eta)\sinh(\l-\m)} \ .\cr}}

\equation{Kcbcbxxx}{h(\l,\m) = 1- i (\l-\m)\ ,\quad
t(\l,\m)= -\frac{1}{(\l-\m)(\l-\m+i)}\ .}

\equation{Kcbad}{\eqalign{K_N\left(\matrix{
\{\l^C\}&\{\l^B\}\cr\{\l^A\}&\{\l^D\}\cr}\right) &= \left(\prod_{j\in
AC}\prod_{k\in DC} f(\l_j^{AC}, \l_k^{DC})\right)\left(\prod_{l\in
AB}\prod_{m\in DB} f(\l_l^{AB}, \l_m^{DB})\right)\cr
&\times\ K_n\left(\matrix{
\{\l^{AB}\}&\{\l^{DC}\}\cr\{\l^{AB}\}&\{\l^{DC}\}\cr}\right)
K_{N-n}\left(\matrix{
\{\l^{AC}\}&\{\l^{DB}\}\cr\{\l^{AC}\}&\{\l^{DB}\}\cr}\right) \ .\cr}}

\equation{hcoeff}{S_N\bigg|_{\nu_j = \l^C_j+i\eta} = K_N\left(\matrix{
\{\l^C\}&\{\l^B\}\cr\{\l^C\}&\{\l^B\}\cr}\right) \prod_{j,k}
\sinh(\l_j^C-\l_k^C-2i\eta)\prod_{m,l}\sinh(\l_m^B-\l_l^C)\ .}

\equation{Sn2}{\eqalign{S_N &= \prod_{j>k}
g(\l_j^C,\l_k^C)g(\l_k^B,\l_j^B) \sum_{}{\rm sgn}(P_C){\rm
sgn}(P_B)\prod_{j,k} h(\l_j^{AB},\l_k^{DC})
\prod_{l,m} h(\l_l^{AC},\l_m^{DB})\cr
&\times\ \prod_{l,k} h(\l_l^{AC},\l_k^{DC})\prod_{j,m}h(\l_j^{AB},\l_m^{DB})
{\rm det}(M^{AB}_{DC}){\rm det}(M^{AC}_{DB})\ ,\cr}}

\equation{detAB}{{\rm det}(A+B) = \sum_{}{\rm sgn}(P_r){\rm sgn}(P_c)
{\rm det}(A_{P_rP_c}){\rm det}(B_{P_rP_c})\ .}

\equation{DQF}{\eqalign{\Phi_A(\l) &= Q_A(\l) + P_D(\l),\quad
\Phi_D(\l) = Q_D(\l) + P_A(\l),\cr
[P_D(\l), Q_D(\m)]&= \ln(h(\l,\m)),\quad [P_A(\l), Q_A(\m)]=
\ln(h(\m,\l))\ .\cr}}

\equation{dfs}{P_a(\l)\dv =0\ ,\quad \ddv Q_a(\l) =0\ ,\ a=A,D\ , \ddv
0)=1\ .}

\equation{dfsnew}{p(\l)\dv =0\ ,\quad \ddv q(\l) =0\ ,\ \ddv 0)=1\ .}

\equation{Sn3}{\eqalign{S_N &= \prod_{j>k} g(\l_j^C,\l_k^C)
g(\l_k^B,\l_j^B) \ddv \det S\dv\ ,\cr
S_{jk} &= t(\l_j^C,\l_k^B) a(\l_j^C)d(\l_k^B)
\exp\left(\Phi_A(\l_j^C)+\Phi_D(\l_k^B) \right)\cr
&\qquad + t(\l_k^B,\l_j^C) d(\l_j^C)a(\l_k^B)
\exp\left(\Phi_D(\l_j^C)+\Phi_A(\l_k^B) \right)\ .\cr}}

\equation{p1}{\eqalign{\ddv \det S\dv\ &=
\sum_{}{\rm sgn}(P_C){\rm sgn}(P_B){\rm det}(M^{AB}_{DC}){\rm
det}(M^{AC}_{DB})\cr
&\times\ddv \exp\left(\sum_{j=1}^n\Phi_A(\l_j^{AC})+\Phi_D(\l_j^{DB})+
\sum_{k=1}^{N-n}\Phi_A(\l_k^{AB})+\Phi_D(\l_k^{DC}) \right)\dv\ .\cr}}

\equation{ndv}{\nddv = \ddv \exp\left(\sum_{j=1}^N
P_D(\l_j^C)+P_A(\l_j^B)\right) \ ,\ \ddv 0)=1\ ,}

\equation{phi}{\eqalign{\varphi(\l) &= p(\l) + q(\l),\ q(\l) =
Q_A(\l)-Q_D(\l) - \nddv Q_A(\l)-Q_D(\l)\dv,\cr
p(\l) &= P_D(\l)-P_A(\l) \ ,\ \nddv q(\l) = 0 = p(\l) \dv\ ,\cr
[p(\l), q(\m)]&= -\ln(h(\l,\m)h(\m,\l))\ ,\ [p(\l),p(\m)]= 0
=[q(\l),q(\m)]\ , [\varphi(\l),\varphi(\m)]=0 \ .\cr }}

\equation{Sn3new}{\eqalign{S_N &= \prod_{j>k} g(\l_j^C,\l_k^C)
g(\l_k^B,\l_j^B) \prod_{j=1}^N
a(\l_j^C)d(\l_j^B)\prod_{j,k}h(\l_j^C,\l_k^B) \nddv \det S\dv\ ,\cr
S_{jk} &= t(\l_j^C,\l_k^B) + t(\l_k^B,\l_j^C) {r(\l_k^B)\over r(\l_j^C)}
\exp\left(\varphi(\l_k^B)-\varphi(\l_j^C) \right)\cr
&\qquad\qquad\times \prod_{m=1}^N {h(\l_k^B,\l_m^B)
h(\l_m^C,\l_j^C)\over h(\l_m^C,\l_k^B) h(\l_j^C,\l_m^B)} ,\cr}}

\equation{bae2}{r(\l_k)\prod_{\scriptstyle j=1\atop \scriptstyle j\neq
k}^N {h(\lambda_k,\lambda_j)\over h(\lambda_j,\lambda_k)} =
(-1)^{N-1}\ , k=1,\ldots ,N\ .}

\equation{od}{S_{jk} = t(\l_j,\l_k) + t(\l_k,\l_j)
\exp\left(\varphi(\l_k)-\varphi(\l_j) \right) \ ,\ j\neq k\ .}

\equation{d}{S_{jj} = i\sin(2\eta)
{\partial\over\partial\l}\left[\ln(r(\l)) +
\sum_{n=1}^N\ln({h(\l,\l_n)\over h(\l_n,\l)})\right]\bigg|_{\l=\l_j} -
2\cos(2\eta) +
i\sin(2\eta){\partial\varphi\over\partial\l}\bigg|_{\l=\l_j}\ .}

\equation{norm2}{\eqalign{&\sigma_N={\langle 0| \prod_{j=1}^{N} C(\lambda_j)
\prod_{k=1}^{N}B(\lambda_k)|0\rangle}= \prod_{j\neq k} f(\l_j,\l_k)
\prod_{j=1}^N a(\l_j)d(\l_j)\nddv \det {\cal N}\dv\ ,\cr
{\cal N}_{jk}&= t(\l_j,\l_k) + t(\l_k,\l_j)
\exp\left(\varphi(\l_k)-\varphi(\l_j) \right) \cr
&\qquad\qquad +i  \sin(2\eta)\delta_{jk}
{\partial\over\partial\l}\left[\ln(r(\l)) +
\sum_{n=1}^N\ln({h(\l,\l_n)\over h(\l_n,\l)})\right]_{\l=\l_j}\ ,\cr}}

\equation{norm3}{\eqalign{&{\langle 0| \prod_{j=1}^{N} C(\lambda_j)
\prod_{k=1}^{N}B(\lambda_k)|0\rangle}= \prod_{j\neq k} f(\l_j,\l_k)
\prod_{j=1}^N a(\l_j)d(\l_j) \det {\cal N^\prime}\ ,\cr}}

\equation{norm3xxz}{
\eqalign{{\cal N^\prime}_{jk} & = \sin(2\eta)\left(
       -K(\l_j,\l_k) + i\ \delta_{jk}
      {\partial\over\partial\l_j}\left[\ln(r(\l_j)) +
      \sum_{n=1}^N\ln({h(\l_j,\l_n)\over h(\l_n,\l_j)})\right]\right) \cr
  & = \sin(2\eta) \left(-K(\l_j,\l_k) + \delta_{jk}
       \left[ i\ {\partial\over\partial\l_j} \ln(r(\l_j)) +
        \sum_{n=1}^N K(\l_j,\l_n) \right] \right) ,}
}

\equation{norm3xxx}{{\cal N^\prime}_{jk}= -\frac{2}{(\l_j - \l_k)^2+1}
+\delta_{jk} \left[i\ {\partial\over\partial\l_j}\ln(r(\l_j)) +
\sum_{n=1}^N \frac{2}{1+(\l_j-\l_n)^2}\right]\ .}

\equation{genfu3}{\eqalign{F(\alpha ,m) &= {1\over \sigma_{N}}
\prod_{j>k}g(\l_j^C,\l_k^C)g(\l_k^B,\l_j^B) \sum_{}\sgn(P_C)\sgn(P_B)
\prod_{I_B,II_B}h(\l^B_I,\l^B_{II})\cr
&\times\prod_{I_C,II_C}h(\l^C_{II},\l^C_{I})
\ddv \det_n s_1(\{\l^C_I\},\{\l^B_I\})\det_{{N}-n}
s_2(\{\l^C_{II}\},\{\l^B_{II}\})\dv \ ,\cr
(s_1(\{\l^C\},\{\l^B\}))_{jk}&= \exp(\alpha) d_2(\l_j^C) a_2(\l_k^B)
\left({\bar s}_1(\{\l^C\},\{\l^B\})\right)_{jk}\ ,\cr
(s_2(\{\l^C\},\{\l^B\}))_{jk}&=  a_1(\l_j^C) d_1(\l_k^B)
\left({\bar s}_2(\{\l^C\},\{\l^B\})\right)_{jk}\ ,\cr}}

\equation{genfu4}{\eqalign{
\left({\bar s}_\gamma(\{\l^C\},\{\l^B\})\right)_{jk} &=
t(\l_j^C,\l_k^B) a_\gamma(\l_j^C)d_\gamma(\l_k^B)
\exp\left(\Phi_{A_\gamma}(\l_j^C)+\Phi_{D_\gamma}(\l_k^B) \right)\cr
&\qquad + t(\l_k^B,\l_j^C) d_\gamma(\l_j^C)a_\gamma(\l_k^B)
\exp\left(\Phi_{D_\gamma}(\l_j^C)+\Phi_{A_\gamma}(\l_k^B) \right)\ .\cr}}

\equation{df}{\eqalign{\Phi_{A_\gamma}(\l) &= Q_{A_\gamma}(\l) +
P_{D_\gamma}(\l),\quad \Phi_{D_\gamma}(\l) = Q_{D_\gamma}(\l) +
P_{A_\gamma}(\l),\cr
[P_{D_\gamma}(\l), Q_{D_\beta}(\m)]&=\delta_{\gamma\beta}
\ln(h(\l,\m)) , \quad [P_{A_\gamma}(\l), Q_{A_\beta}(\m)]=
\delta_{\gamma\beta} \ln(h(\m,\l))\ .\cr}}

\equation{df2}{P_a(\l)\dv =0\ ,\quad \ddv Q_a(\l) =0\ ,\ a=A_\gamma
,D_\gamma\ , \ddv 0)=1\ .}

\equation{genfu5}{\eqalign{F(\alpha ,m) &= {1\over \sigma_{N}}
\prod_{j>k}g(\l_j^C,\l_k^C)g(\l_k^B,\l_j^B)\ddv \det{\cal M}\dv\ ,\cr
{\cal M}_{jk}&=\left({s}_1(\{\l^C\},\{\l^B\})\right)_{jk}
\exp\left(\psi_{D_1}(\l^C_j)+\psi_{A_1}(\l^B_k)\right) \cr
&\qquad\qquad +\left({s}_2(\{\l^C\},\{\l^B\})\right)_{jk}
\exp\left(\psi_{A_2}(\l^C_j)+\psi_{D_2}(\l^B_k)\right)\ , \cr}}

\equation{genfu6}{\eqalign{F(\alpha ,m) &= \frac{\nddv \det{{\cal
G}}\dv}{\det {\cal N}^\prime}\ ,\cr
{{\cal G}}_{jk}&= t(\l_j,\l_k) + t(\l_k,\l_j) {r_1(\l_j)\over
r_1(\l_k)}\exp\left(\varphi_2(\l_k) - \varphi_2(\l_j)\right) \cr
&+ \exp\left(\alpha +\varphi_4(\l_k)-\varphi_3(\l_j)\right)
\left[t(\l_k,\l_j) + t(\l_j,\l_k) {r_1(\l_j)\over
r_1(\l_k)}\exp\left(\varphi_1(\l_j) - \varphi_1(\l_k)\right)\right]\cr
&+ \delta_{jk}\omega\left(L D(\l_j) +\sum_n K(\l_j,\l_n)\right),\cr}}

\equation{ndf3}{\eqalign{
\varphi_a(\lambda) &= p_a(\lambda) + q_a(\lambda)\ ,\
\nddv q_a(\l) = 0 = p_a(\l) \dv \ , \nddv 0)=1\ ,\ a=1\ldots 4\ , \cr
 [q_b(\mu), p_a(\lambda)] &= \left(\matrix{1&0&1&0\cr 0&1&0&1\cr
0&1&1&1\cr 1&0&1&1\cr}\right) \ln(h(\lambda ,\mu)) +
\left(\matrix{1&0&0&1\cr 0&1&1&0\cr 1&0&1&1\cr
0&1&1&1\cr}\right) \ln(h(\mu, \lambda)),\ a,b=1\ldots 4  \ .\cr}}

\equation{ndf}{\eqalign{\phi_1(\l)&= \Phi_{A_1}(\l)-\Phi_{D_1}(\l)\ ,\
\phi_2(\l)= \Phi_{A_2}(\l)-\Phi_{D_2}(\l)\ ,\cr
\phi_3(\l)&=
\psi_{A_2}(\l)-\psi_{D_1}(\l)-\Phi_{D_1}(\l)+\Phi_{A_2}(\l)\ ,\cr
\phi_4(\l)&=
\psi_{A_1}(\l)-\psi_{D_2}(\l)+\Phi_{A_1}(\l)-\Phi_{D_2}(\l)\ .\cr}}

\equation{ndv2}{\nddv = \ddv \exp\left(\sum_{j=1}^{N}
P_{D_2}(\l_j)+P_{A_2}(\l_j)+{\cal P}_{D_1}(\l_j)+{\cal
P}_{A_1}(\l_j)\right) \ ,\ \nddv 0)=1\ ,}

\equation{ndf2}{\eqalign{\varphi_a(\l) &= \phi_a(\l) -\nddv
\phi_a(\l)\dv = p_a(\l)+q_a(\l)\ ,\ a=1,\ldots 4\ .\cr}}

\equation{genfu7}{\eqalign{&F(\alpha ,m) = {1\over \sigma_{N}}
\prod_{j>k}f(\l_j,\l_k)f(\l_k,\l_j)\prod_{j=1}^{N} a(\l_j)d(\l_j)\nddv
\det{\bar{\cal M}}\dv\ ,\cr
&{\bar{\cal M}}_{jk}= t(\l_j,\l_k) + t(\l_k,\l_j) {r_2(\l_k)\over
r_2(\l_j)}\exp\left(\phi_2(\l_k) - \phi_2(\l_j)\right) \cr
&+ {r(\l_k)\over r(\l_j)}\exp\left(\alpha
+\phi_4(\l_k)-\phi_3(\l_j)\right) \left[t(\l_k,\l_j) +
t(\l_j,\l_k) {r_1(\l_j)\over  r_1(\l_k)}\exp\left(\phi_1(\l_j) -
\phi_1(\l_k)\right)\right] .\cr}}

\equation{bae3}{\eqalign{r_2(\l_k)\prod_{\scriptstyle j=1\atop
\scriptstyle j\neq k}^{N} {h(\lambda_k,\lambda_j)\over
h(\lambda_j,\lambda_k)} &= {(-1)^{{N}-1}\over r_1(\l_k)}\ ,\cr
{1\over r_2(\l_k)}\prod_{\scriptstyle j=1\atop
\scriptstyle j\neq k}^{N} {h(\lambda_j,\lambda_k)\over
h(\lambda_k,\lambda_j)} &= (-1)^{{N}-1} r_1(\l_k)\ , k=1,\ldots
,{N}\ \cr}}

\equation{genfu8}{\eqalign{&F(\alpha ,m) = {1\over \sigma_{N}}
\prod_{j\neq k}f(\l_j,\l_k)\prod_{j=1}^{N} a(\l_j)d(\l_j)\nddv
\det{{\cal G}}\dv\ ,\cr
&{{\cal G}}_{jk}= t(\l_j,\l_k) + t(\l_k,\l_j) {r_2(\l_k)\over
r_2(\l_j)}e^{\varphi_2(\l_k) - \varphi_2(\l_j)}e^{\kappa_2(\l_k) -
\kappa_2(\l_j)}\cr &+ {r(\l_k)\over r(\l_j)}e^{\alpha
+\varphi_4(\l_k)-\varphi_3(\l_j)} e^{\kappa_4(\l_k) - \kappa_3(\l_j)}
\left[t(\l_k,\l_j) + t(\l_j,\l_k) {r_1(\l_j)\over
r_1(\l_k)}\exp\left(\varphi_1(\l_j) - \varphi_1(\l_k)\right)\right]
.\cr}}

\equation{J}{ \det J = \prod_{j=1}^{N} 2\pi L\rho(\l_j)\ .}

\equation{V}{\eqalign{
V (\l, \m)  &=  \frac{-\sin(2\eta)}{\sinh(\l-\m)} \bigg\{ {1\over
\sinh(\l-\m + 2i\eta)} - {e_2^{-1} (\l) e_2(\m) \over\sinh(\m - \l +
2i\eta)} \cr
&\quad + \exp(\alpha +\varphi_4(\mu) - \varphi_3(\lambda))\bigg(
{-1\over\sinh(\mu-\lambda + 2i\eta)} + {e_1^{-1} (\mu) e_1(\lambda)
\over\sinh(\lambda - \mu + 2i\eta)}\bigg)\bigg\}\ ,  \cr}}

\equation{e}{e_2 (\l) = \bigg( {\sinh(\l + i\eta) \over \sinh(\l - i\eta)}
\bigg)^{m} \exp(\varphi_2 (\l ))
\qquad e_1 (\l) = \bigg( {\sinh(\l - i\eta) \over \sinh(\l + i\eta)}
\bigg)^{m} \exp(\varphi_1 (\l )) \ .}

\equation{genfu9}{F(\alpha ,m)= {\nddv\det (id\ +{1\over 2
\pi}\widehat{V})\dv\over \det (id\ - {1 \over 2\pi} \widehat{K})} \ .}

\equation{genfuxxx}{\eqalign{F_{XXX}(\alpha ,m)= {(0\vert\det
(1+{1\over 2 \pi}\widehat{V}_{XXX})\vert 0) \over \det (1- {1 \over 2\pi}
\widehat{K}_{XXX})} \ ,\cr }}

\equation{Vxxx}{\eqalign{
V (\l, \m)  &=  \frac{1}{\l-\m} \bigg\{ {1\over
\l-\m + i} - {e_2^{-1} (\l) e_2(\m) \over \m - \l +i} \cr
&\quad + \exp(\alpha +\varphi_4(\mu) - \varphi_3(\lambda))\bigg(
{-1\over\mu-\lambda + i} + {e_1^{-1} (\mu) e_1(\lambda)
\over\lambda - \mu + i}\bigg)\bigg\}  \cr}}

\equation{U}{U(\l,\m) = + {i \left( 1 - e^\alpha \right) \over
\sinh(\l-\m)}\left\{ 1 - \left( {{e^{2\l}+i}\over{e^{2\l}-i}}\
{{e^{2\m}-i}\over{e^{2\m}+i}} \right)^{m} \right\}\ .}

\equation{vxx}{\eqalign{&F_{XX}(\alpha ,m)= {(0\vert\det (id+{1\over 2
\pi}\widehat{V})\vert 0) } \ ,\cr &V_{XX}(\l, \m)  =  \frac{2i}{\sinh
2(\l-\m)} \bigg\{ 1 - e_2^{-1}(\l) e_2(\m)-  e^{\alpha +\varphi_3(\mu)
- \varphi_3(\lambda)}\bigg(1- e_1^{-1} (\mu)
e_1(\lambda)\bigg)\bigg\}\ ,\cr}}

\def\sgn{{\rm sgn}}
\def\p{p^{^{\!\!\!\circ}}}
\def\q{q^{^{\!\!\!\circ}}}
\def\dv{|0)}
\def\ddv{(0|}
\def\nddv{{(\tilde 0|}}
\def\qed{{\smallfonts\bf q.e.d.}}
\def\vac{|0\rangle}
\def\dvac{\langle 0|}
\def\dbvac{\bar{\langle 0|}}
\def\bvac{\bar{\vac}}
\def\up{\uparrow}
\def\half{{1\over 2}}
\def\l{\lambda}
\def\m{\mu}
\def\eps{\epsilon}
\def\half{{1\over 2}}
\def\frac#1#2{{#1\over #2}}
\def\B#1{{\cal B}(#1)}
\def\D#1{{\cal D}(#1)}
\def\C#1{{\cal C}(#1)}
\def\A#1{{\cal A}(#1)}

%
%
\pagenumstyle{blank}
\sectionnumstyle{arabic}
\footnoteskip=2pt
\footnotenumstyle{arabic}
\line{\smallfonts\it June 1994\hfil BONN-TH-94-14}
\line{\smallfonts\it \hfil ITP-UH-09/94}
\line{\smallfonts\it \hfil ITP-SB-94-24}
\vskip2em
\baselineskip=32pt
\begin{center}
{\bigsize{\sc Determinant Representation for Correlation Functions\\
of Spin-1/2 XXX and XXZ Heisenberg Magnets}}
\end{center}
\vfil
\baselineskip=12pt
\vskip 2em
\begin{center}
{\bigsize
Fabian H.L. E\sharps ler\footnote[$\ \flat$]{\sc e-mail:
fabman@avzw02.physik.uni-bonn.de}}\vskip 0.3cm
{\it Physikalisches Institut der Universit\"at Bonn\vskip 3pt
Nussallee 12, 53115 Bonn, Germany}
{\bigsize
\vskip .5cm
Holger Frahm\footnote[$\ \natural$]{\sc e-mail:
frahm@itp.uni-hannover.de}\vskip .3cm}
{\it Institut f\"ur Theoretische Physik, Universit\"at
  Hannover\vskip3pt
D-30167 Hannover, Germany}
{\bigsize\vskip .5cm
Anatoli G. Izergin\footnote[$\ \clubsuit$]{\sc e-mail:
izergin@lomi.spb.su}\vskip .3cm}
{\it Department of Mathematical Institute of Sciences, Academy of
Sciences\vskip 3pt of Russia, St. Petersburg,  POMI, SU-199106,
Fontanka 27, Russia}
{\bigsize
\vskip .5cm
Vladimir E. Korepin\footnote[$\ \sharp$]{\sc e-mail:
korepin@max.physics.sunysb.edu}\vskip .3cm}
{\it Institute for Theoretical Physics\vskip 3pt
SUNY at Stony Brook, Stony Brook, NY~~11794-3840, USA}

\end{center}

\vfil
\vskip 2em
\centertext{\bfs \bigsize ABSTRACT}
\vskip\belowsectionskip
\begin{narrow}[4em]
\noindent
\baselineskip=16pt
We consider correlation functions of the spin-$\half$ XXX and XXZ
Heisenberg chains in a magnetic field. Starting from the algebraic
Bethe Ansatz we derive representations for various correlation
functions in terms of determinants of Fredholm integral operators.

\end{narrow}
\vfil

\break

\pagenumstyle{arabic}
\baselineskip=16pt
\sectionnum=0
{\sc\section{Introduction}} Despite the great advances made over the last
sixty years in the study of integrable quantum models, evaluation of their
correlation functions still poses a formidable problem. Quite recently
there has been significant progress in this direction: the group at RIMS
succeeded in deriving integral representations for some correlation
functions of the Heisenberg XXZ model defined by the hamiltonian (1.1) for
$\Delta >1$ by taking advantage of the infinite quantum affine symmetry of
the model on the infinite chain\upref rims1, rims2/. The isotropic (XXX)
limit $\Delta\rightarrow 1$ was obtained in [\putref{rims3,kieu}]. These
integral representations are most powerful for studying the {\sl short
distance} asymptotics of correlators, whereas it is not obvious how to
extract the long-distance behaviour. Also it is not straightforward to
extend this approach to the critical regime $-1\leq \Delta < 1$ or to
include an external magnetic field.\\
Precisely these issues can be very naturally addressed in the framework of
a different approach to studying correlation functions in integrable
models, which was carried out in
[\putref{k1,k2,iik1,iik2,iik3,iiks,iikv,ks1,ks2}] for the example of the
Bose gas\footnote{A detailed and complete exhibition of this
work can be found in the book [\putref{vladb}]}.  We call this method the
{\sl Dual Field Approach} (DFA). The DFA is directly based on the
(algebraic) Bethe-Ansatz solution of the model and thus is applicable
to a large variety of correlation functions and integrable models. It
allows to derive
explicit expressions for the {\sl large distance} asymptotics of
correlation functions (even at finite temperature), and the inclusion of an
external magnetic field poses no problem. The DFA thus nicely complements
the approach of the RIMS group. In a series of papers we will apply the DFA
to the Heisenberg XXZ and XXX chains at zero temperature in a magnetic
field $h$, {\sl i.e.} the hamiltonian
$$\putequation{H}$$
where $\sigma^z=\left(\matrix{1&0\cr 0&-1\cr}\right)$,
$\sigma^x=\left(\matrix{0&1\cr 1&0\cr}\right)$,
$\sigma^y=\left(\matrix{0&-i\cr i&0\cr}\right)$, and
$\Delta=\cos(2\eta)$.\vskip .5cm\noindent

There are four main steps in the DFA: First the model needs to be
``solved'' by means of the Algebraic Bethe Ansatz. Then one uses this
solution to express correlation functions in terms of determinants of
Fredholm integral operators. In step three these determinants are embedded
in systems of integrable integro-difference equations (IDE).  Finally the
large-distance asymptotics of the correlators is extracted from a
Riemann-Hilbert problem for the IDE's. As the computations for the various
steps are rather involved we will only deal with the first two steps here,
{\sl i.e.} review the known Bethe Ansatz solution for the XXZ and XXX
chains and then derive determinant representations for correlation
functions. In two following publications we will present steps
three\upref IDE/ and four.

{\sc\section{A Short Review of Algebraic Bethe Ansatz}}
Let us review a few main features of the Algebraic Bethe Ansatz (ABA)
for both XXZ and XXX Heisenberg magnets, in order to fix notations for
things to come.
The XXX case can of course be obtained by taking a certain limit of the XXZ
case, but in practice this is more difficult than treating the XXX case
separately from the beginning. Thus we will treat both cases on an equal
footing throughout this paper.

Starting point and central object of the Quantum Inverse Scattering Method
is the R-matrix, which is a solution of the Yang-Baxter equation. For the
case of the XXZ and XXX models it is of the form
$$\putequation{R}$$
where for XXZ
$$\putequation{fgxxz}$$
and for the XXX-case
$$\putequation{fgxxx}$$
The R-matrix is a linear operator on the tensor product of two
two-dimensional linear spaces: $R(\mu)\in
End(\Complex^2\otimes\Complex^2)$.  {}From the R-matrix (\putlab{R}) one
can construct an $L$-operator of a "fundamental spin model"\upref vladb/ by
considering the matrix $R(\mu)\Pi$, where $\Pi$ is the permutation matrix
on $\Complex^2\otimes\Complex^2$, and then making it into an
operator-valued matrix by identifying one of the linear spaces with the
two-dimensional Hilbert space ${\cal H}_n$ of $SU(2)$-spins over the n'th
site of a lattice of length $L$
$$\putequation{lop}$$
The Yang-Baxter equation for $R$ implies the following intertwining
relations for the $L$-operator
$$\putequation{intL}$$
{}From the ultralocal $L$-operator the monodromy matrix is constructed as
$$\putequation{monodr}$$
The intertwiner (\putlab{intL}) can be lifted to the level of the monodromy
matrix
$$\putequation{intT}$$
Below we will repeatedly use especially the following matrix elements
of (\putlab{intT})
$$\putequation{intTme}$$
By tracing (\putlab{intT}) over the matrix space one then finds that the
{\sl transfer matrices} $\tau(\mu)=tr(T(\mu))=A(\mu)+D(\mu)$ commute for any
values of spectral parameter $\mu$, {\sl i.e.}
$[\tau(\mu),\tau(\nu)]=0$. {}From this it follows that the transfer matrix
is the generating functional of an infinite number of mutually commuting
conserved quantum operators (via expansion in powers of spectral
parameter). One of these operators is the hamiltonian
$$\putequation{Hxxz}$$
Below we also make use of some properties of {\sl inhomogenous} XXX and XXZ
models, which are constructed in the following way: we first note that the
intertwiner for the $L$-operator (\putlab{intL}) still holds, if we shift
both spectral parameters $\l$ and $\mu$ by an arbitrary amount $\nu_n$,
{\sl i.e.}
$$\putequation{intL2}$$
The reason for this fact is of course that the $R$-matrix only depends on
the difference of spectral parameters. We now can construct a monodromy
matrix as
$$
   \putequation{Tinhom}
$$

The ABA deals with the construction of simultaneous eigenstates of the
transfer matrix and the hamiltonian. Starting point is the choice of a
{\sl reference state}, which is a trivial eigenstate of $\tau(\mu)$. In
our case we make the choice $\vac = |\up\up\up\ldots\up\rangle =
\otimes_{n=1}^L |\up\rangle_n$, {\sl i.e.} we choose the completely
ferromagnetic state. The action of the $L$-operator (\putlab{lop}) on
$|\up\rangle_n$ can be easily computed and implies the following actions of
the matrix elements of the monodromy matrix for the XXZ case
$$
   \putequation{ADxxz}
$$
whereas in the XXX case
$$
   \putequation{adxxx}
$$
{}From (\putlab{ADxxz}) it follows that $B(\lambda)$ plays the role of a
creation operator, {\sl i.e.} a Fock space of states can be constructed as
$$
   \putequation{states}
$$
The requirement that the states (\putlab{states}) ought to be eigenstates
of the transfer matrix $\tau(\mu)$ puts constraints on the allowed values
of the parameters $\lambda_n$: the set $\{\lambda_j\}$ must be a solution
of the following system of coupled algebraic equations, called {\sl Bethe
equations}
$$
   \putequation{bae}
$$
These equations are the basis for studying ground state, excitation
spectrum and thermodynamics of Bethe Ansatz solvable models. For the case
of the XXZ model with $\Delta > -1$ (the case we are interested in here) it
was proved by C.N. Yang and C.P. Yang in [\putref{yaya1, yaya2}] that the
ground state is characterized by a set of {\sl real} $\lambda_j$ subject to
the Bethe equations (\putlab{bae}). Without an external magnetic field
($h=0$) their number is $N=L/2$. In the thermodynamic limit the ground
state is described by means of an integral equation for the density of
spectral parameters $\rho(\lambda)$
$$
   \putequation{gsie}
$$
where the integral kernel $K$ and the driving term $D$ are given by
$$
   \putequation{kernel}
$$
For the XXX case we have
$$
   \putequation{kernelXXX}
$$
Here $\Lambda$ depends on the external magnetic field $h$.  The physical
picture of the ground state is that of a filled Fermi sea with boundaries
$\pm \Lambda$. The dressed energy of a particle in the sea is given by the
solution of the integral equation
$$
   \putequation{gsie2}
$$
The requirement of the vanishing of the dressed energy at the Fermi
boundary $\eps(\pm\Lambda)=0$ determines the dependence of $\Lambda$ on
$h$. For small $h$ this relation can be found explicitly by means of a
Wiener-Hopf analysis\upref vladb/.
For $h\geq h_c=(2\cos\eta)^2$ the system is in the saturated
ferromagnetic state, which corresponds to $\Lambda=0$.

{\sc\section{Two-Site Generalized Model}}
For the evaluation of correlation functions the so-called ``two-site
generalized model'' has proven an extremely useful tool. From the
mathematical point of view this is simply the application of the
co-product associated with the algebra defined by (\putlab{intT}). The
main idea is to divide the chain of length $L$ into two parts and
associate a monodromy matrix with both sub-chains, {\sl i.e.}
$$\putequation{t12}$$
In terms of $L$-operators the monodromy matrices are given by
$$\putequation{t12l}$$
By construction it is clear that both monodromy matrices $T(i,\mu)$
fulfill the same intertwining relation (\putlab{intT}) as the complete
monodromy matrix $T(\mu)$.
Similarly the reference state for the complete chain is decomposed
into a direct product of reference states $\vac_i$ for the two
sub-chains $\vac = \vac_2\otimes\vac_1$. The resulting structure can
be summarized as
$$\putequation{abcdi}$$
where the eigenvalues $a$ and $d$ in (\putlab{ADxxz}) are given by
$a(\mu) =a_2(\mu)a_1(\mu)$ and $d(\mu) =d_2(\mu)d_1(\mu)$.
The creation operators $B(\mu)$ for the complete chain are decomposed
as $B(\mu) = A_2(\mu)\otimes B_1(\mu) +B_2(\mu) \otimes D_1(\mu)$,
which implies that eigenstates of the transfer matrix can be
represented as
$$\putequation{bbb}$$
where the sum is over all partitions
$\{\lambda_j^I\}\cup\{\lambda_k^{II}\}$ of the set $\{\lambda_j\}$ with
${\rm card}\{\lambda^I\}=n_1$, ${\rm card}\{\lambda^{II}\}=n_2=N-n_1$.  A
similar equation holds for dual states
$$\putequation{ccc}$$

{\sc\section{Reduction of Correlators to Scalar Products}}
In this section we reduce the problem of evaluating correlators of the
form $\langle\sigma_j^z\sigma_k^z\rangle$ (where $\langle\rangle$
denotes the normalized zero temperature vacuum expectation value, {\sl
i.e.} the expectation value with respect to the antiferromagnetic
ground state described by (\putlab{gsie})-(\putlab{gsie2})) to
the computation of certain scalar products between states given by the
Algebraic Bethe Ansatz. We start by noting that due to translational
invariance it is sufficient to consider the correlator $G(m)=\langle
\sigma_m^z\sigma_1^z\rangle$. In terms of the operators $q_j=\half
(1-\sigma_j^z)$ the correlator takes the form
$$\putequation{g}$$
where we have again used translational invariance. The quantity
$\langle q_1\rangle$ is nothing but the density of down spins in the
ground state and can thus be reexpressed as $\langle q_1\rangle =
\int_{-\Lambda}^{\Lambda} d\Lambda\rho(\Lambda)$. The first term in
(\putlab{g}) is expressed in terms of the quantity $Q_1(m)=\sum_{j=1}^m
q_j= \sum_{j=1}^m \sigma_j^-\sigma_j^+$ as follows
$$ \langle q_mq_1\rangle = \half {\widehat\Delta }\langle
(Q_1(m))^2\rangle\ ,$$
where ${\widehat\Delta}$ (not to be confused with the inhomogeneity
$\Delta$ in the XXZ hamiltonian) is the lattice laplacian
${\widehat\Delta} f(j) = f(j)+f(j-2)-2f(j-1)$. Putting everything
together we obtain
$$\putequation{szsz}$$
The only nontrivial quantity to determine is thus $\langle
(Q_1(m))^2\rangle = {\partial^2\over\partial\alpha^2}\langle\exp(\alpha
Q_1(m))\rangle\bigg|_{\alpha=0}$. We will now use the two-site
generalized model to express the ``generating functional''
$$\putequation{genfu}$$
in terms of scalar products: we take the first sub-chain to contain
sites $1$ to $m$ and the second one sites $m+1$ to $L$. We note that
$Q_1(m)$ now acts only on the first sub-chain and simply counts the
number of down spins. Using (\putlab{bbb}) and (\putlab{ccc}) in
(\putlab{genfu}) we obtain
$$\putequation{genfu2}$$
where the sum is over all partitions
$$\{\l^B_{I_B}\}\cup \{\l^B_{II_B}\} = \{\l\},\ \{\l^B_{I_B}\}\cap
\{\l^B_{II_B}\} = \emptyset\ ,\
\{\l^C_{I_C}\}\cup \{\l^C_{II_C}\} = \{\l\},\ \{\l^C_{I_C}\}\cap
\{\l^C_{II_C}\} = \emptyset$$
of the set $\{\lambda\}$ with ${\rm card}\{\lambda_{I_B}\}={\rm
  card}\{\lambda_{I_C}\}= n_1$, ${\rm   card}\{\lambda_{II_C}\}= {\rm
  card}\{\lambda_{II_B}\}= N-n_1$ and
$$\putequation{Sn}$$
Note that due to
(\putlab{abcdi}) and (\putlab{intT}) (for $B_i(\mu), C_i(\mu)$) we only
need to consider partitions such that the size of partitions $I_B$ and
$I_C$ (and $II_B$ and $II_C$) are the same. In the following section
we will show that scalar products of the form appearing in
(\putlab{genfu2}) can be expressed as determinants and then use this
fact to obtain a determinant representation for $F(\alpha ,m)$.

A particularly simple correlator to compute within this approach is
the ``Emptiness Formation Probability''. This correlation function is
defined as
$$ P (m) =  \langle\prod^m_{j=1} \half (\sigma_j^z + 1) \rangle\quad
,$$
and physically corresponds to the probability to find a string of
ferromagnetically ordered adjacent spins in the (antiferromagnetic)
ground state\upref vladb/. It can be obtained from $F(\alpha , m)$ in
the limit $\alpha\to -\infty$
$$ P(m) = \lim_{\alpha\to -\infty} F(\alpha ,m)\ .$$

{\sc\section{Scalar Products}}
We now turn to the investigation of scalar products of the form
$$\putequation{scpr}$$
Here we {\sl do not} assume that the sets of spectral parameters
$\{\l^B\}$ and $\{\l^C\}$ are the same, and we also {\sl do not}
impose the Bethe equations (\putlab{bae}), because out goal is to
determine the scalar products occurring in (\putlab{genfu2}). The {\sl
  norms}\upref k1/ are a special case of (\putlab{scpr}).
{}From (\putlab{intTme}) and (\putlab{ADxxz}) it follows
that scalar products can be represented as
$$\putequation{Ks}$$
where the sum is over all partitions of $\{\l^C\}\cup \{\l^B\}$ into
two sets $\{\l^A\}$ and $\{\l^D\}$. The coefficients $K_N$ are
functions of the $\l_j$ and are completely determined by the
intertwiner (\putlab{intT}). In particular the $K_N$'s are identical
for the homogeneous model (\putlab{monodr}) and the {\sl
inhomogeneous} model (\putlab{Tinhom}), {\sl i.e.} the $K_N$'s are
independent of the inhomogeneities $\{\nu_n\}$ and also do not depend
on the lattice length $L$ as long as $N<L$. We will exploit this fact
by considering special inhomogeneous models for which all terms but one
in the sum in (\putlab{Ks}) vanish, and then represent this term as a
determinant. The basic tool for representing scalar products as
determinants is a Theorem due to Izergin, Coker and Korepin\upref
i1,ick/, which deals with determinant representations for the {\sl
  partition functions} of inhomogeneous XXZ and XXX models
constructed according to (\putlab{Tinhom}):

\vskip .3cm
\noindent{\bfs Theorem 1:} \sl Consider an inhomogeneous XXZ chain of
even length $N$ with inhomogeneities $\nu_j,\ j=1\ldots N$. Let $\vac$
and ${\bar{\vac}}$ be the ferromagnetic reference states with all spins
up and down respectively. Let $\B{\mu}$ and $\C{\mu}$ be the
  creation/annihilation operators over the reference state $\vac$.
  Then the following determinant representations hold for the XXZ magnet
$$\putequation{detZ1}$$
where
$$\putequation{detZ2}$$
A similar representation holds for the XXX magnet.
\vskip .5cm\rm
Let us now derive explicit expressions for the coefficients $K_N$. It
will be convenient to work with the following sets of spectral
parameters
$$ \eqalign{\{\l^{AC}\} &= \{\l^A\}\cap \{\l^{C}\},\
\{\l^{DC}\} = \{\l^D\}\cap \{\l^{C}\},\cr
\{\l^{AB}\} &= \{\l^A\}\cap \{\l^{B}\},\
\{\l^{DB}\} = \{\l^D\}\cap \{\l^{B}\},\cr}$$
with cardinalities
$$\eqalign{n&={\rm card}\{\l^{DC}\}={\rm card}\{\l^{AB}\}\ ,\cr
N-n&={\rm card}\{\l^{AC}\}={\rm card}\{\l^{DB}\}\ .\cr}$$
The partition with $n=0$ is characterized by $\{\l^{AC}\}=\{\l^{C}\}$,
$\{\l^{DB}\}=\{\l^{B}\}$, $\{\l^{AB}\}=\emptyset = \{\l^{DC}\}$. The
corresponding coefficient $K_N\left(\matrix{
  \{\l^C\}&\{\l^B\}\cr\{\l^C\}&\{\l^B\}\cr}\right)$ is called {\sl
  highest coefficient}.

\vskip .3cm
\noindent{\bfs Lemma 1:} {\sl For highest coefficients the following
  determinant representation holds}
$$\putequation{Kcbcb}$$
{\sl For the XXZ magnet we find}
$$\putequation{Kcbcbxxz}$$
{\sl and in the XXX case we have instead}
$$\putequation{Kcbcbxxx}$$
\noindent{\bfs Proof:}
We will carry out the proof for the XXZ case, the XXX case is similar.
Consider an inhomogeneous XXZ model on a lattice of length $N$ with
inhomogeneities $\nu_j = \l_j^C+i\eta$. We have $a(\l) =
\prod_{j=1}^N\sinh(\l-\l_j^C-2i\eta)$ and $d(\l) =
\prod_{j=1}^N\sinh(\l-\l_j^C)$. Inspection of (\putlab{Ks}) yields that in
this situation only one term in the sum of the r.h.s of (\putlab{Ks})
survives, namely the one with $\{\l^D\} = \{\l^B\}$.  Thus for this special
scalar product we obtain
$$\putequation{hcoeff}$$
On the other hand $B(\l)$ flips one spin, and as we have chosen $N$ to
be the length of the lattice we find that $\prod_{j=1}^NB(\l_j)\vac$
is proportional to the ferromagnetic state with all spins flipped, and
thus orthogonal to all states in a basis other than ${\bar{\vac}}$.
Thus
$$S_N\bigg|_{\nu_j = \l^C_j+i\eta} = {\langle 0| \prod_{j=1}^{N}
  C(\lambda_j^C)\bvac \dbvac \prod_{k=1}^{N}B(\lambda_k^B)|0\rangle}\ .$$
By Theorem 1 both factors can be represented as determinants. By direct
computation we find for one of the factors
$$ \langle 0| \prod_{j=1}^{N}   C(\lambda_j^C)\bvac  =
  \prod_{j,k}\sinh(\l_k^C -\l_j^C-2i\eta) \ .$$
Using the determinant representation given by Theorem 1 on the other
factor we arrive at (\putlab{Kcbcb}).\quad \qed
\vskip .5cm
\noindent{\bfs Lemma 2:} {\sl Arbitrary coefficients $K_N$ are
expressed in terms of highest coefficients as follows}
$$\putequation{Kcbad}$$
\noindent{\bfs Proof:}
We again will only treat the XXZ case explicitly, the XXX case being
very similar. Consider an inhomogeneous XXZ model with inhomogeneities
$\{\nu_j\} = \{\l_j^{AB}+i\eta\}\cup \{\l_j^{AC}+i\eta\}$. Now
only the term proportional to $K_N\left(\matrix{
\{\l^C\}&\{\l^B\}\cr\{\l^A\}&\{\l^D\}\cr}\right)$ in the sum on the
r.h.s of. (\putlab{Ks}) survives. Proceeding as in the proof of Lemma
1 above we arrive at (\putlab{Kcbad}).\quad \qed
\vskip .5cm
Combining the results of Lemmas 1 and 2 with (\putlab{Ks}) we arrive
at the following expression for general scalar products of XXZ and XXX
magnets
$$\putequation{Sn2}$$
where $P_C$ is the permutation $\{ \l_1^{AC},\ldots ,
\l_n^{AC},\l_1^{DC},\ldots ,\l^{DC}_{N-n}\}$ of $\{\l_1^C,\ldots
,\l_N^C\}$, $P_B$ is the permutation $\{ \l_1^{DB},\ldots ,
\l_n^{DB},\l_1^{AB},\ldots ,\l^{AB}_{N-n}\}$ of $\{\l_1^B,\ldots
,\l_N^B\}$, $sgn(P)$ is the sign of the permutation $P$, and
$$ \eqalign{\left(M^{AB}_{DC}\right)_{jk} &= t(\l_j^{AB},\l_k^{DC})
  d(\l_k^{DC})a(\l_j^{AB}) ,\quad t(\l,\m) = {(g(\l,\m))^2\over
    f(\l,\m)}\ .\cr}$$
Note that (\putlab{Sn2}) is formally the same as the corresponding
expression for the delta-function Bose gas\upref vladb/, only the
functions $f(\l,\m)$, $g(\l,\m)$ (and thus also $h$ and $t$), $a(\l)$
and $d(\l)$ are different.

{\sc\section{Dual Fields}}
The most important step in the DFA follows next: we introduce {\sl
  dual quantum fields} in order to simplify (\putlab{Sn2}) and obtain
a manageable expression for scalar products. This step was first
carried out in [\putref{k2}] for the delta-function Bose gas. The
XXX and XXZ cases of interest here can be treated very similarly, so
that we will be brief in our discussion.
The fundamental observation is that the r.h.s. in(\putlab{Sn2}) looks
like the determinant of the {\sl sum} of two matrices:\vskip .3cm

\noindent{\bfs Lemma 3:} {\sl Let $A$ and $B$ be two $N\times N$
  matrices over $\Complex$. Then the determinant of their sum can be
  decomposed as follows}
$$\putequation{detAB}$$
{\sl Here $P_r$ and $P_c$ are partitions of the $N$ rows and columns
  into two subsets ${\cal R}$, ${\bar{\cal R}}$ and ${\cal C}$
  $\bar{\cal C}$ of cardinalities $n$ (for ${\cal R}$, ${\cal C}$) and
  $N-n$ (for $\bar{\cal R}$, $\bar{\cal C}$) respectively,
  $A_{P_rP_c}$ is the $n\times n$ matrix obtained from $A$ by removing
  all $\bar{\cal R}$-rows and $\bar{\cal C}$-columns,
  and $B_{P_rP_c}$ is the $N-n\times N-n$ matrix obtained from $B$ by
  removing all ${\cal R}$-rows and ${\cal C}$-columns.
  Finally ${\rm sgn}(P_r)$ is the parity of the permutation obtained
  from $(1,\ldots ,N)$ by moving all ${\cal R}$-rows to the
  front.}\vskip .3cm
\noindent{\bfs Proof:} See [\putref{vladb}] p.221 ff.
\vskip .5cm
Comparison of (\putlab{Sn2}) with Lemma 3 shows that one does not get the
$h(\l,\m)$-factors by simply taking the determinant of the sum of the
matrices $M_{jk}$. This leads to the introduction of two {\sl dual quantum
fields} $\Phi_A(\l)$ and $\Phi_D(\l)$ which are represented as sums of
``momenta'' $P_A$ and ``coordinates'' $Q_A$ as follows
$$\putequation{DQF}$$
All other commutators of $P$'s and $Q$'s vanish. A very important
property of the fields $\Phi$ is that they commute for different
values of spectral parameters
$$ [\Phi_A(\l), \Phi_D(\m)] = 0 =[\Phi_A(\l), \Phi_A(\m)]
=[\Phi_D(\l), \Phi_D(\m)]\ .$$
The dual quantum fields act on a bosonic Fock space with reference
states $\dv$ and $\ddv$ defined by
$$\putequation{dfs}$$
Using the dual fields it is now possible to recast (\putlab{Sn2}) as a
determinant of the sum of two matrices
\vskip .3cm
\noindent{\bfs Theorem 2:} {\sl Scalar products for the Heisenberg XXZ
  and XXX magnets can be represented as determinants in the following way}
$$\putequation{Sn3}$$
\noindent{\bfs Proof:} Using Lemma 3 to expand the determinant in
(\putlab{Sn3})we arrive at
$$\putequation{p1}$$
Evaluating the expectation value of the dual quantum fields by means
of (\putlab{DQF}) and (\putlab{dfs}) we arrive at (\putlab{Sn2}).\ \qed
\vskip .5cm
It is possible to further simplify (\putlab{Sn3}) by eliminating one
dual field: we define a new dual vacuum $\nddv$ according to
$$\putequation{ndv}$$
and a new dual field
$$\putequation{phi}$$
In terms of this field we obtain the following determinant
representation
$$\putequation{Sn3new}$$
where $r(\l) = {a(\l)\over d(\l)}$. Now we have all the machinery
ready to tackle the problem of representing (\putlab{genfu2}) as a
determinant.

{\sc\section{On Norms}}
In this section we will have a closer look at norms of Bethe wave
functions. These were evaluated for both XXZ and XXX models in
[\putref{k1}], so that the answers are already known. Here we will
consider norms as special cases of scalar products in order to build
up some machinery needed below for further analysis of (\putlab{Sn3}).
We will treat the XXZ case in detail and quote the results for XXX. In
order to study norms we ought to set $\{\l^C\}=\{\l^B\}$ in
(\putlab{Sn3}) and then impose the Bethe equations (\putlab{bae}).
Immediately some problems arise as the diagonal elements of the matrix
$S$ in (\putlab{Sn3}) become ill-defined (``$0\over 0$'') and
have to be investigated more carefully. The off-diagonal matrix
elements are easily dealt with. The Bethe equations (\putlab{bae})
together with the antisymmetry property $g(\l,\m)=-g(\m,\l)$ imply
$$\putequation{bae2}$$
Thus we obtain
$$\putequation{od}$$
To obtain the diagonal matrix elements we take the limit
$\l_j\rightarrow \l_k$ in the matrix $S$ and use l'Hospital's rule
(here we have to make use of the explicit expressions for the
functions $f$, $g$, $a$, $d$, {\sl etc} for the XXZ case)
$$\putequation{d}$$
To obtain this expression we also have made use of the Bethe equations
(\putlab{bae2}). We observe that the last two terms in (\putlab{d})
are precisely what one obtains when taking the limit
$\l_j\rightarrow\l_k$ in (\putlab{od}). Putting everything together we
find the following expression for the norm (\putlab{Sn})
$$\putequation{norm2}$$
where we now interpret the first two terms in $\cal N$ in the sense of
l'Hospital for the diagonal elements. For the case of the XXX magnet we
have to replace $\sin(2\eta)$ by $1$ and use the functions $f,g,h,t$
following from (\putlab{fgxxx}). There is one further simplification:
it was shown in [\putref{k1}] that the expectation value of the dual
field part in (\putlab{norm2}) is such that the dual fields can be
simply set equal to zero, {\sl i.e.} we can replace $\nddv {\rm det}{\cal
  N}\dv$ by ${\rm det}{\cal N}^\prime$, where ${\cal N}^\prime$ is
obtained from $\cal N$ by dropping the $\exp(\varphi)$-terms. Then a
further simplification takes place as
$$t(\l_j,\l_k) +t(\l_k,\l_j) = -\sin(2\eta) K(\l_j,\l_k), $$
where $K$ is defined in (\putlab{kernel}). This is
summarized in the following theorem due to Korepin\upref k1/
\vskip .5cm
\noindent{\bfs Theorem 3:} {\sl Norms for the Heisenberg XXZ
  and XXX magnets can be represented as determinants in the following way}
$$\putequation{norm3}$$
{\sl For the XXZ case the matrix ${\cal N}^\prime$ is given by}
$$\putequation{norm3xxz}$$
{\sl where $K(\l,\m)$ and $h(\l,\m)$ are defined in (\putlab{kernel})
and (\putlab{Kcbcbxxz}) respectively. For the XXX case we have instead}
$$\putequation{norm3xxx}$$
\vskip .3cm
\noindent{\bfs Proof:} see [\putref{k1}]\ .

{\sc\section{Correlators on the Finite Chain}}
Let us now come back to the generating functional for correlators
(\putlab{genfu2}). We will now use the machinery built up in the last
few sections to express $F(\alpha,m)$ as a determinant. We will
proceed in two steps: we first will analyse (\putlab{genfu2}) {\sl
  without} using that $\{\l^B\}=\{\l^C\}=\{\l\}$ and {\sl without}
imposing the Bethe-equations (\putlab{bae}). In the second step we
will then impose these two constraints.
Using (\putlab{Sn3new}) we can represent the scalar products in the
two-site generalized models in (\putlab{genfu2}) as determinants
\vskip .5cm
\noindent{\bfs Lemma 4:}
$$\putequation{genfu3}$$
{\sl where}
$$\putequation{genfu4}$$
{\sl Here the dual fields are defined according to}
$$\putequation{df}$$
{\sl All other commutators vanish. The reference state $\dv$ and its
  dual $\ddv$ are annihilated by all momenta/coordinates respectively}
 $$\putequation{df2}$$
\vskip .5cm
\noindent{\bfs Proof:} We use (\putlab{Sn3}) to express both
${_1}\dvac\prod_{I_C} C_1(\l^C_{I_C}) \prod_{{I_B}}
B_1(\l^B_{I_B})\vac_1\ $ \hfill\break and ${_2}\dvac\prod_{II_C}
C_2(\l^C_{II_C})
\prod_{II_B} B_2(\l^B_{II_B})\vac_2\ $ as determinants. We are led to
introduce two sets of dual fields (one for each scalar product)
$\Phi_{A_\gamma}(\l),\ \Phi_{D_\gamma}(\l),\ \gamma =1,2$ with
commutation relations given by (\putlab{df}). The two kinds of dual
fields are completely independent of each other (all commutators
bewteen momenta/coordinates of different sets vanish). The
representation (\putlab{genfu3})-(\putlab{genfu4}) is now obtained by
direct computation, where the $\sgn(P_B)\sgn(P_C)$ arises upon
taking the factor $\prod_{j>k}g(\l_j^C,\l_k^C)g(\l_k^B,\l_j^B)$ in
front of the sum due to $g(\l,\m)=-g(\m,\l)$.\ \qed
\vskip .3cm
We now observe that (\putlab{genfu3}) is basically of the same
structure as (\putlab{Sn2}). Thus, in analogy with (\putlab{Sn3}), we
can introduce new dual quantum fields and reexpress $F(\alpha ,m)$ as
{\sl a single} determinant.
\vskip .5cm
\noindent{\bfs Lemma 5:} {\sl Consider the set of four commuting dual
  quantum fields }
$$ \eqalign{\psi_{D_1}(\l)&= {\cal Q}_{D_1}(\l)+{\cal P}_{A_2}(\l)\ ,
  \psi_{A_1}(\l)= {\cal Q}_{A_1}(\l)+{\cal P}_{D_2}(\l)\ ,\cr
  \psi_{D_2}(\l)&= {\cal Q}_{D_2}(\l)+{\cal P}_{A_1}(\l)\ ,
  \psi_{A_2}(\l)= {\cal Q}_{A_2}(\l)+{\cal P}_{D_1}(\l)\ ,\cr}$$
{\sl with commutation relations of the momenta/coordinates given by}
$$[{\cal P}_{D_\gamma}(\l), {\cal Q}_{D_\beta}(\m)]=\delta_{\gamma\beta}
\ln(h(\l,\m)) , \quad [{\cal P}_{A_\gamma}(\l), {\cal Q}_{A_\beta}(\m)]=
\delta_{\gamma\beta} \ln(h(\m,\l))\ .$$
{\sl All other commutators vanish. The action of the dual fields on
  the dual reference states is given by ${\cal P}_a(\l)\dv=0$,
  $\ddv{\cal Q}_a(\l)=0$, $a=A_1,A_2,D_1,D_2$. Then the following
  determinant   representation holds}
$$\putequation{genfu5}$$
{\sl where $({s}_\gamma)_{jk}$ are given by (\putlab{genfu3}).}
\vskip .3cm
\noindent{\bfs Proof:} The proof is analogous to the one for Theorem
2, only the expectation value of dual quantum fields is slightly
different. \qed
\vskip .5cm
So far we have not used the fact that we are dealing with expectation
values of Bethe states, {\sl i.e.} we have neither used the fact that
$\{\l^C\}=\{\l^B\}=\{\l\}$ nor imposed the Bethe equations
(\putlab{bae}). In the next step we will impose these constraints. The
discussion will be reminiscent of section $7$ above. The result is
summarized in the following\vskip .3cm

\noindent{\bfs Theorem 4:} {\sl The generating funtional $F(\alpha
  ,m)$ can be represented as a ratio of determinants in the following way}
$$
   \putequation{genfu6}
$$
{\sl where $r_1(\l) = {a_1(\l)/ d_1(\l)}$, $K(\l,\m)$ and $D(\l)$
  are defined in (\putlab{kernel}) (\putlab{kernelXXX}),
  $\omega=\sin(2\eta)$ for XXZ and $\omega =-1$ for XXX, and the
  commuting dual fields $\varphi_a$ are defined according to}
$$\putequation{ndf3}$$
{\sl Here all terms not proportional to $\delta_{jk}$ in ${\cal
    G}_{jk}$ are understood in the sense of l'Hospital for the diagonal
  elements.} \vskip .3cm
\noindent{\bfs Proof:}
We start by defining a new dual vacuum $\nddv$
and a new set of dual fields according to
$$\putequation{ndv2}$$
$$\putequation{ndf}$$
The fields $\phi_a(\l)$ can be decomposed into momenta $\p_a$ and
coordinates $\q_a$ (by using (\putlab{df}) and the definitions of $\psi_a$
given in Lemma 5), which are found to obey the commutation relations
(\putlab{ndf3}). By straighforward rewriting of (\putlab{genfu5}) in terms
of the new fields and the new dual reference state we obtain
$$
\putequation{genfu7}
$$
Here we have used that
$$
   \ddv \exp\left(\sum_{j=1}^{N} \phi_{A_2}(\l_j)+\phi_{D_2}(\l_j)+
   \Phi_{A_2}(\l_j)+\Phi_{D_2}(\l_j)\right) =
   \prod_{j,k}h(\l_j,\l_k)\nddv\ .
$$
It is found that whereas ${\p}_a(\l)\dv =0$, the coordinates
${\q}_a(\l)$ of $\phi_a(\l)$ do not annihilate the new dual reference
state $\nddv$. Therefore we ``shift'' $\phi_a(\l)$ by subtracting
their vacuum expectation values in analogy with (\putlab{phi})
$$
\putequation{ndf2}
$$
By construction the $p$'s and $q$'s have the same commutations
relations (\putlab{ndf3}) as the momenta/coordinates ${\p}_a(\l)$
and ${\q}_a(\l)$ of the $\phi_a(\l)$'s. Furthermore $p_a(\l)\dv=0$
and $\nddv q_a(\l)=0$ for $a=1\ldots 4$. The shifts are found to be
$$ \kappa_a(\l)=\nddv \phi_a(\l)\dv = (1-\delta_{a1})
\sum_j\ln\left({h(\l,\l_j)\over h(\l_j,\l)}\right)\ .$$
If we replace the fields $\phi_a$ in (\putlab{genfu7}) by the fields
$\varphi_a$ we pick up additional factors due to the shifts
$$
   \putequation{genfu8}
$$
The off-diagonal matrix elements of $G$ can be further simplified by simply
imposing the Bethe equations.  Rewriting the Bethe equations (\putlab{bae})
as
$$
   \putequation{bae3}
$$
we find that the additional factors take the form
$$
   \eqalign{\exp\left(\kappa_2(\l_k)-\kappa_2(\l_j)\right){r_2(\l_k)\over
    r_2(\l_j)}&= {r_1(\l_j)\over r_1(\l_k)}\ ,\cr
   \exp\left(\kappa_4(\l_k)-\kappa_3(\l_j)\right){r(\l_k)\over r(\l_j)}&=
   1\ .\cr}
$$
Inserting this into (\putlab{genfu7}) we arrive at (\putlab{genfu6}) {\sl
without} the term proportional to $\delta_{jk}$, {\sl i.e.} we have proved
(\putlab{genfu6}) for the off-diagonal matrix elements. To get the diagonal
matrix elements we have to investigate the limit $\l_j\rightarrow\l_k$ of
(\putlab{genfu8}) in detail. In the limit $\l_j\rightarrow\l_k$ the sum of
the first two terms in $G_{jk}$ and the expression in brackets are both of
the form ``${0\over 0}$''. By using l'Hospital's rule we find analogously
to section $7$ above
$$
\eqalign{&\lim_{\l_j\rightarrow\l_k}\left(t(\l_j,\l_k) + t(\l_k,\l_j)
{r_2(\l_k)\over r_2(\l_j)}e^{\varphi_2(\l_k) -
  \varphi_2(\l_j)}e^{\kappa_2(\l_k) -\kappa_2(\l_j)}\right) =\cr
&=-2\cosh(2i\eta) + \sinh(2i\eta){\partial \varphi_2(\l)\over\partial\l}
\bigg|_{\l=\l_j}+\sinh(2i\eta) {\partial\over\partial\l_j}\left[
\ln(r_2(\l_j))+\sum_n\ln({h(\l_j,\l_n)\over h(\l_n,\l_j)})\right] ,\cr}$$
$$\eqalign{\lim_{\l_j\rightarrow\l_k}&\left(t(\l_j,\l_k) + t(\l_k,\l_j)
{r_1(\l_j)\over r_1(\l_k)}e^{\varphi_2(\l_k) - \varphi_2(\l_j)}\right)=\cr
&=-2\cosh(2i\eta) + \sinh(2i\eta)\left({\partial \varphi_2(\l_j)\over
\partial\l_j} +m \frac{\sinh(2i\eta)}{\sinh(\l_j+i\eta)\sinh(\l_j-i\eta)}
\right)\ .\cr}
$$
Using these expressions we find that the diagonal terms of $G$ in
(\putlab{genfu8}) are equal to the diagonal terms of $G$ in
(\putlab{genfu6}) if we keep in mind that the first two lines of $G_{jk}$
in (\putlab{genfu6}) are interpreted a la l'Hospital for $j=k$.  Last but
not least we insert the expression (\putlab{norm3}) for $\sigma_{N}$ in the
resulting expression and arrive at (\putlab{genfu6}).  This completes the
proof of the theorem.\ \qed
\vskip .5cm
Theorem 4 states the determinant representation for the generating
functional $F(\alpha ,m)$ on a {\sl finite} chain of length $L$. As always
in Bethe Ansatz solvable models significant simplifications take place if
we take the thermodynamic limit $L\rightarrow\infty$.  This is done in the
next section.

{\sc\section{Thermodynamic Limit}}
The results of taking the thermodynamic limit of (\putlab{genfu6}) and main
results of this paper are summarized in the following
\vskip .3cm

\noindent{\bfs Theorem 5:} {\sl In the thermodynamic limit the
  generating funtional $F(\alpha ,m)$ for the case of the XXZ magnet
  can be represented as a ratio of determinants of Fredholm integral
  operators $\left(id + {1\over 2\pi}\widehat{V}\right)$ and
  $\left(id\ -{1\over 2\pi}\widehat{K}\right)$ in the following way}
$$
   \putequation{genfu9}
$$
{\sl Here $\nddv$ and $\dv$ are the vacua of the dual bosonic Fock space
  defined in (\putlab{ndf3}) and the integral operators act on functions
  $f$ defined on the interval $[-\Lambda, \Lambda ]$ according to}
$$
\eqalign{(id -{1\over 2\pi}\widehat{K})*f\bigg|_\l &=
f(\l)-\frac{1}{2\pi}\int_{-\Lambda}^\Lambda d\mu K(\l,\m) f(\m)\ ,\cr
(id +{1\over 2\pi}\widehat{V})*f\bigg|_\l &=
f(\l)+\frac{1}{2\pi}\int_{-\Lambda}^\Lambda d\mu V(\l,\m) f(\m)\
,\cr}
$$
{\sl where the kernel $K(\l,\m)$ is defined in (\putlab{kernel}) and the
  kernel of $\widehat V$ is given by}
$$
   \putequation{V}
$$
$$
   \putequation{e}
$$
{\sl The dual fields $\varphi_a(\l)$ are defined in (\putlab{ndf3}),
   with $h(\l,\m)$ given in (\putlab{Kcbcbxxz}).}
\vskip .3cm
\noindent{\bfs Proof:}
We begin by taking the thermodynamic limit for the norm
$\sigma_{N}$ (\putlab{norm3}). We first write ${\cal N}^\prime$
as the product of two matrices:
$$
   {{\cal N}^\prime}_{jk} =  \sin(2\eta)
         \sum_m I_{jm} J_{mk}, \quad
   I_{jm} = \delta_{jm} - \frac{K_{jm}}{\theta_m},\quad
   J_{jm} = \delta_{jm}\theta_m,
$$
where $\theta_m = LD(\l_m)+\sum_n K(\l_m,\l_n)$. Here $D$ and $K$ are
defined in (\putlab{gsie})--(\putlab{kernel}). The determinant of ${\cal
N}^\prime$ is the product of the determinants of $I$ and $J$.  Next we use
that the set of roots $\{\l_j\}$ describes the ground state and the roots
thus obey the equations
$$
   2\pi L\rho(\l_j) -\sum_{k=1}^{N} K(\l_j,\l_k) = L D(\l_j)\
,\qquad   j=1\ldots {N}\ ,
$$
which is the discrete version of (\putlab{gsie}). Here $\rho(\l_j) =
\frac{1}{L(\l_{j+1}-\l_j)}$, which becomes $\rho(\l)$ defined by
(\putlab{gsie}) in the thermodynamic limit. We thus can rewrite
$\theta_m=2\pi L\rho(\l_m)$, which leads to
$$
   \putequation{J}
$$
In the thermodynamic limit the matrix $I$ turns into an integral operator
${\widehat I}=id -{1\over 2\pi}{\widehat K}$
$$
   {\widehat I}*f\bigg|_\l = f(\l)-\frac{1}{2\pi}\int_{-\Lambda}^\Lambda
   d\mu K(\l,\m) f(\m)\ ,
$$
where $K$ is the kernel of $\widehat K$ defined by (\putlab{kernel}).
\vskip .3cm
The matrix $G_{jk}$ in (\putlab{genfu6}) is treated in a very similar
way. We rewrite it as a product
$$
   G_{jk} = \sin(2\eta) \sum_m W_{jm} J_{mk}\ ,
$$
where $J_{jm}=\delta_{jm}2\pi L\rho(\l_m)$ is the same as above, and
$$
 \eqalign{
   W_{jk}&= \delta_{jk} + {1\over\sin(2\eta)\theta_k} \bigg\{
   t(\l_j,\l_k) + t(\l_k,\l_j) {r_1(\l_j)\over
   r_1(\l_k)}\exp\left(\varphi_2(\l_k) - \varphi_2(\l_j)\right) \cr
   &+ \exp\left(\alpha +\varphi_4(\l_k)-\varphi_3(\l_j)\right)
   \left[t(\l_k,\l_j) + t(\l_j,\l_k) {r_1(\l_j)\over
   r_1(\l_k)}\exp\left(\varphi_1(\l_j) -
   \varphi_1(\l_k)\right)\right]\bigg\}\ .\cr}
$$
In the thermodynamic limit the matrix $W_{jk}$ turns into an integral
operator ${\widehat W}= id + {1\over 2\pi}{\widehat V}$, with kernel
$V(\l,\m)$ defined by (\putlab{V}). Thus we obtain (\putlab{genfu9}) in the
thermodynamic limit.\ \qed

\vskip .5cm

For the XXX chain a determinant representation is obtained in an analogous
way. The result is found to be
$$
   \putequation{genfuxxx}
$$
where ${\widehat K}_{XXX}$ and ${\widehat V}_{XXX}$ are integral operators
with kernels $K(\lambda,\mu)$ from (\putlab{kernelXXX}) and
$$
   \putequation{Vxxx}
$$
$$
   e_2 (\l) = \bigg( {\l + \frac{i}{2} \over \l -\frac{i}{2}}
   \bigg)^{m} \exp(\varphi_2 (\l )) \ , \qquad
   e_1 (\l) = \bigg( {\l - \frac{i}{2} \over \l + \frac{i}{2}}
   \bigg)^{m} \exp(\varphi_1 (\l ))\ .
$$
The dual fields $\varphi_a(\l)$ are again defined in (\putlab{ndf3}), but
now $h(\l,\m) = 1- i (\l-\m)$.

The Emptiness Formation Probability can be easily obtained from
(\putlab{genfu9}) and (\putlab{genfuxxx}) by setting $\alpha=
-\infty$, which corresponds to dropping the second line in the
expressions for the kernel of $\widehat V$ in (\putlab{V}) and
(\putlab{Vxxx}). For the XXX case this exactly reproduces the result
of [\putref{kieu}].

{\sc\section{Some limiting Cases}}

It is quite straightforward to evaluate the determinants in
(\putlab{genfu9}) for strong magnetic fields $h\sim
h_c=(2\cos\eta)^2$, in which the ground state is very close to the
ferromagnetic vacuum and $\Lambda \ll 1$.
The near asymptotics ($m\ll (\pi/2\Lambda) \tan\eta$) of the EFP for
the XXZ case follows to be
$$P(m)= 1 - \left({2\Lambda\over \pi\sin\eta}\cos\eta\right) m\ .$$
Using (\putlab{gsie}) and (\putlab{gsie2}) this reproduces the obvious
result
$$P(m)= 1 - \half\left(1 -\langle \sigma_j^z\rangle\right) m=
1-{m\over\pi}\sqrt{h_c-h}\ ,\ h\to h_c,\ h<h_c\ .$$

Another interesting limiting case (which allows to make contact with
known results) is the $XX0$ free fermionic limit of the $XXZ$ model,
where $\eta = {3\pi\over 4}$ in (\putlab{V}) and (\putlab{kernel}).
This case has been previously considered in [\putref{cikt,iiks2}],
where a determinant representation for $F(\alpha ,m)$ was found, which
does not involve dual quantum fields. Taking the free fermionic limit
of (\putlab{ndf3}) we obtain
$$
   [q_b(\mu), p_a(\lambda)] = \left(\matrix{2&0&1&1\cr 0&2&1&1\cr
   1&1&2&2\cr 1&1&2&2\cr}\right) \ln(\cosh(\lambda -\mu))\ .
$$
Thus we can choose $\varphi_3(\lambda) = \varphi_4(\lambda)$ and
reduce the number of dual fields to $3$. Furthermore we have
$\sin(4\eta)=0$ and therefore $K(\lambda ,\mu)=0$. The determinant
formula then reads
$$
\putequation{vxx}
$$
where now
$$
   e_2(\l) = \bigg( {{e^{2\l}-i}\over{e^{2\l}+i}} \bigg)^{m}
      \exp(\varphi_2 (\l ))\ ,\qquad
   e_1(\l) = \bigg( {{e^{2\l}+i}\over{e^{2\l}-i}} \bigg)^{m}
      \exp(\varphi_1 (\l ))\ .
$$
This expression has to be compared with the result obtained in
[\putref{cikt}] which reads after transforming their expression to the
notation used in the present paper
$$
   F_{XX}(\alpha,m) = \det\left( id + {1\over2\pi} \widehat{U}\right)\ .
$$
Here $\widehat{U}$ is the integral operator defined in terms of the
kernel
$$
\putequation{U}
$$
The proof of the equivalence of (\putlab{U}) and (\putlab{vxx}) is
quite tedious and might be presented elsewhere.

{\sc\section{Discussion}}
In this paper we have derived determinant representations for
spin-spin correlation functions in XXZ and XXX Heisenberg models.
Other correlation functions can be treated similarly.
In a following publication we will use the determinant representation
to embed the correlation functions in systems of integrable
integro-difference equations\upref IDE/.

\vskip .3cm\noindent
Finally we would like to emphasize that the Dual Field Method can be
applied to any correlation function in any integrable model.

\vskip .3cm
\centerline{\sc Acknowledgements:}
\vskip .2cm
H.F.\ acknowledges partial support by the Deutsche Forschungsgemeinschaft
under Grant No.\ Fr~737/2--1.
V.E.K.\ was supported by the National Science Foundation under
Grant No.\ 9321165.

\begin{putreferences}
\centerline{{\sc References}}
\smallfonts
\vskip .5cm
\reference{cikt}{F. Colomo, A.G. Izergin, V.E. Korepin, V. Tognetti,\
\PLA{169}{1992}{243}.}
\reference{rims1}{B. Davies, O. Foda, M. Jimbo, T. Miwa, A.
Nakayashiki,\ \CMP{151}{1993}{89}.}
\reference{IDE}{F.H.L. E\char'31ler, H. Frahm, A.R. Its, V.E.
Korepin,\ {\sl in preparation}.}
\reference{iik1}{A.R. Its, A.G. Izergin, V.E. Korepin,\
\CMP{130}{1990}{471}.}
\reference{iik2}{A.R. Its, A.G. Izergin, V.E. Korepin,\
\Phy{D53}{1991}{187}.}
\reference{iik3}{A.R. Its, A.G. Izergin, V.E. Korepin,\
\PLA{141}{1989}{121}.}
\reference{iiks}{A.R. Its, A.G. Izergin, V.E. Korepin, N. Slavnov,\
\IJMPB{4}{1990}{1003}.}
\reference{iiks2}{A.R. Its, A.G. Izergin, V.E. Korepin, N. Slavnov,\
\PRL{70}{1993}{1704}.}
\reference{iikv}{A.R. Its, A.G. Izergin, V.E. Korepin, G.G. Varzugin,\
\Phy{D54}{1992}{351}.}
\reference{i1}{A.G. Izergin,\ \SPD{32}{1987}{878}.}
\reference{ick}{A.G. Izergin, D. Coker, V.E. Korepin,\
\JPA{25}{1992}{4315}.}
\reference{ik1}{A.G. Izergin, V.E. Korepin,\ \CMP{113}{1987}{117}.}
\reference{rims2}{M. Jimbo, K. Miki, T. Miwa, A. Nakayashiki,\
\PLA{168}{1992}{256}.}
\reference{k1}{V.E. Korepin,\ \CMP{86}{1982}{391}.}
\reference{k2}{V.E. Korepin,\ \CMP{113}{1987}{177}.}
\reference{vladb}{V.E. Korepin, G. Izergin and N.M. Bogoliubov,\ {\sl
Quantum Inverse Scattering Method, Correlation Functions and Algebraic
Bethe Ansatz}, Cambridge University Press, 1993}
\reference{kieu}{V.E. Korepin, A.G. Izergin, F.H.L. E\char'31ler, D.
Uglov,\ \PLA{190}{1994}{182}.}
\reference{ks1}{V.E. Korepin, N. Slavnov,\ \CMP{129}{1990}{103}.}
\reference{ks2}{V.E. Korepin, N. Slavnov,\ \CMP{136}{1991}{633}.}
\reference{rims3}{A. Nakayashiki,\ {\sl preprint}.}
\reference{yaya1}{C.N. Yang, C.P. Yang,\ \PR{150}{1966}{321}.}
\reference{yaya2}{C.N. Yang, C.P. Yang,\ \PR{150}{1966}{327}.}

\end{putreferences}
\bye